\documentclass[journal,10pt]{IEEEtran}
\usepackage{float,array}
\usepackage{graphicx}
\usepackage{adjustbox}
\usepackage{lineno,hyperref,cite}
\modulolinenumbers[5]
\usepackage{moreverb}
\usepackage{lettrine}
\usepackage{xcolor,colortbl}
\definecolor{Gray}{gray}{0.95}

\definecolor{kugray5}{RGB}{224,224,224}
\usepackage{amssymb,amsmath}
\usepackage[linesnumbered,ruled,vlined]{algorithm2e}
\usepackage{etoolbox}
\newcounter{algoline}

\AtBeginEnvironment{algorithm}{\setcounter{algoline}{0}}
\usepackage{multirow}
\usepackage{caption}
\usepackage{pifont}
\usepackage{breqn}
\usepackage{caption}
\usepackage{subcaption}
\usepackage{mathtools}
\usepackage{stfloats}
\usepackage{amsmath}
\usepackage{pdflscape}
\usepackage{booktabs, caption, makecell}
\usepackage{threeparttable}
\usepackage{multirow}
\usepackage{bigstrut}
\usepackage{mwe}

\usepackage{wrapfig}
\usepackage{textcomp}
\usepackage{adjustbox}
\usepackage[none]{hyphenat}
\usepackage{rotating}
\newcolumntype{L}[1]{>{\raggedright\let\newline\\\arraybackslash\hspace{0pt}}m{#1}}
\newcolumntype{C}[1]{>{\centering\let\newline\\\arraybackslash\hspace{0pt}}m{#1}}
\newcolumntype{R}[1]{>{\raggedleft\let\newline\\\arraybackslash\hspace{0pt}}m{#1}}
\usepackage{bookmark}

\begin{document}

\title{Swarm of UAVs for Network Management in 6G: A Technical Review}

\author{Muhammad~Asghar~Khan~,~Neeraj~Kumar,~\IEEEmembership{Senior Member,~IEEE},~Syed~Agha~Hassnain~Mohsan, ~Wali~Ullah~Khan,~\IEEEmembership{Member,~IEEE},~Moustafa M.~Nasralla,~\IEEEmembership{Senior Member,~IEEE}, Mohammed~H.~Alsharif,\\Justyna Żywiołek,~and~Insaf~Ullah

\thanks{MA. Khan and I. Ullah are with the Hamdard Institute of Engineering and Technology, Islamabad 45550, Pakistan. E-mails: m.asghar@hamdard.edu.pk; insafktk@gmail.com

N.Kumar is with Department of Computer Science and Information Engineering, Asia University, Taiwan and School of Computer Science, University of Petroleum and Energy Studies, Dehradun, Uttarakhand; E-mail: neeraj.kumar.in@ieee.org 

SAH Mohsan is with Optical Communication Laboratory, Ocean College, Zhejiang University, Zheda Road 1, Zhoushan 316021, China; E-mail: hassnainagha@zju.edu.cn

W. U. Khan is with Interdisciplinary Centre for Security, Reliability and Trust (SnT), University of Luxembourg, 1855 Luxembourg City, Luxembourg. E-mail: waliullah.khan@uni.lu

MM.Nasralla is with Communications and Networks Engineering Department, Prince Sultan University, Riyadh, Saudi Arabia;E-mail:mnasralla@psu.edu.sa

MH. Alsharif is with Sejong University, Seoul 05006, South
Korea. E-mail: malsharif@sejong.ac.kr.

J. Żywiołek is with Czestochowa University of Technology, Czestochowa, Poland. E-mail: justyna.zywiolek@wz.pcz.pl
}
}

\markboth{Accepted in IEEE Transactions on Network and Service Management, vol. xx, no. yy, Month 202z}
{}

\maketitle

\begin{abstract}
 Fifth-generation (5G) cellular networks have led to the implementation of beyond 5G (B5G) networks, which are capable of incorporating autonomous services to swarm of unmanned aerial vehicles (UAVs). They provide capacity expansion strategies to address massive connectivity issues and guarantee ultra-high throughput and low latency, especially in extreme or emergency situations where network density, bandwidth, and traffic patterns fluctuate. On the one hand, 6G technology integrates AI/ML, IoT, and blockchain to establish ultra-reliable, intelligent, secure, and ubiquitous UAV networks. 6G networks, on the other hand, rely on new enabling technologies such as air interface and transmission technologies, as well as a unique network design, posing new challenges for the swarm of UAVs.Keeping these challenges in mind, this article focuses on the security and privacy, intelligence, and energy-efficiency issues faced by swarms of UAVs operating in 6G mobile network. In this state-of-the-art review, we integrated blockchain and AI/ML with UAV networks utilizing the 6G ecosystem. The key findings are then presented, and potential research challenges are identified. We conclude the review by shedding light on future research in this emerging field of research.
\end{abstract}
\begin{IEEEkeywords}
UAV; 6G Networks; Security and Privacy; Blockchain; AI/ML; Energy Efficiency.
\end{IEEEkeywords}
\IEEEpeerreviewmaketitle

\section{Introduction}
\label{sec:intro}
While 5G mobile systems are being deployed around the globe, researchers have started to envision 6G networks to integrate the functions of sensing, communication, computation, and control.  Following in the footsteps of the 5G wireless network, 6G wireless network is expected to provide massive connectivity to millions of interconnected devices with diverse quality of service (QoS) requirements, ubiquitous coverage, high degree of embedded artificial intelligence (AI), efficient use of energy, and adaptive network security \cite{1}. In addition, unmanned aerial vehicles (UAVs) can play a significant part in the 6G ecosystem since flying devices are expected to densely occupy aerial space, operating as a network layer between ground and space networks. As a result, UAVs communicate with ground and satellite stations, forming a space-air-ground network that paves the path for fully integrated 6G heterogeneous networks \cite{2,3}.UAVs can be used as a vertical component in a variety of settings, such as aerial base stations, access points (APs), relays, or flying mobile terminals, to improve the coverage, reliability, and energy efficiency of 6G wireless networks.
\begin{table}[h!]
\begin{center}
\caption{List of key acronyms.}
\label{notations}
\begin{tabular}{lp{5.8cm}}
\toprule 
Label & Explanation \\
\midrule
5G & Fifth Generation\\
B5G & Beyond Fifth Generation\\      
6G & Sixth Generation\\
AI & Artificial Intelligence\\
ANN & Artificial Neural Network\\
AP & Access Point\\
BC & BlockChain\\
BS & Base Station\\
CNN & Convolutional Neural Network \\
DoS & Denial-of-Service\\
eMBB & enhanced Mobile BroadBand\\
FSO & Free Space Optics\\
GNSS & Global Navigation Satellite System\\
HAPS & High Altitude Platform System\\
GCS & Ground Control Station\\
GPS & Global Positioning System\\
IDS & Intrusion Detection System\\
IoT & Internet of Things\\
LoS & Line-of-Sight\\
LTE & Long-term Evolution\\
MBS & Macro Base Station\\
MEC & Mobile Edge Computing\\
MIMO & Multiple-Input Multiple-Output\\
ML & Machine Learning\\
mMTC & massive Machine-Type Communications\\
mmWave & millimeter Wave\\
NFV&Network Functions Virtualization\\
NOMA & Non-Orthogonal Multiple Access\\
QBC & Quantum Backscatter Communications\\ 
QoE & Quality of Experience\\
QoS & Quality of Service\\
QML & Quantum Machine Learning\\
PoW & Proof of Work\\
RAN & Radio Access Network\\
RF & Radio Frequency\\
RIS & Reconfigurable Intelligent Surface\\
SBS & Small Base Station\\
SDN & Software-defined Networking\\
THz & TeraHertz\\
UAV & Unmanned Aerial Vehicle\\
uHDD & ultra-High Data Density\\
uHSLLC & ultra-High Speed Low Latency Communications \\
uMUB & ubiquitous Mobile Ultrabroadband\\
URLLC & Ultra-Reliable Low-Latency Communications\\
VLC & Visible-Light Communications\\
XR& Extended Reality\\
WPT& Wireless Power Transfer\\

\bottomrule 
\end{tabular}
\end{center}
\end{table}
Despite the fact that UAVs have been tested for deployment in existing cellular networks and approved by 3GPP for seamless integration with 5G networks, their full potential can only be realised if their communications are extended to the space network, which can be accomplished more efficiently with the help of a 6G network.For future UAV networks, new technical scenarios are emerging. Traditional UAV networks are evolving into enhanced networks, in which UAVs connect with one another in an ad hoc manner. This new breed of UAV networks is made up of evolving nodes capable of executing computing, communication, and control functions, as well as developing self-organizing and self-sustaining capabilities and assuring connectivity.

Towards the fulfillment of this grand vision, 6G anticipates maximum spectral utilization employing multi-band high-spread spectrum, as well as high frequency bands such as Sub-6 GHz, millimeter-wave (mmWave), and terahertz (THz) to support high data throughput transmission links \cite{4,5,6,7,8,9,10}.  Optical wireless communications (OWC), which is also viewed as a key enabling technology for providing high data rates at low energy consumption. To improve the programmability, scalability, and flexibility of a UAV network, novel frameworks based on software defined networking (SDN) and network function virtualization (NFV) can be adopted. It can also make connection switching and programmable metasurfaces easier, as well as simplifying network management. In addition, Integrating network resources with cloud computing and edge computing paradigms will provide low-cost, high-flexibility on-demand computation and storage capabilities \cite{11,12,13,14,15,16,17,18,19,20}.

The evaluation of cellular communication in light of technological advancements for various UAV applications is depicted in Fig.1. Tab.I presents a list of commonly used acronyms in this review, while Tab.II offers a brief comparison of 1G and 6G communications based on some key performance indicators (KPIs). Tab.II demonstrates that 6G networks will provide speeds exceeding 1 Tbps and latency of less than 1ms. Moreover, the integration of AI/ML, blockchain, and edge computing technologies into UAV networks using 6G networks offers several research opportunities.With this integration, the inherent challenges of conventional UAV systems, such as limited processing resources, energy efficiency, privacy, and security, can be overcome. Concerns about security and privacy are rarely addressed in the design of UAVs. UAVs are susceptible to a number of security vulnerabilities due to limited and inadequate on-board computing and energy capabilities. Consequently, 6G should emphasize security and privacy, and the wireless research community should pay special attention to them for UAV communication and networking \cite{21}. Researchers all over the world are proposing artificial intelligence (AI)/machine learning (ML)  \cite{22,23,24,25,26,27}, quantum machine learning (QML) \cite{28}, blockchain \cite{29}, terahertz (THz) communication \cite{30,31,32}, fog/edge computing \cite{33}, visible light communication (VLC) \cite{34}and other cutting-edge technologies as prerequisites for the deployment of 6G wireless networks. To better comprehend the integration of UAV networks with enabling technologies in 6G, see Fig.4.These technologies can help improve the performance of the UAV network in a variety of ways, including QoS, QoE, security, fault management, path planning, navigation, and energy efficiency, all of which are key challenges associated with UAVs.
\begin{figure*}
\centering
    \includegraphics[width=0.9\textwidth]{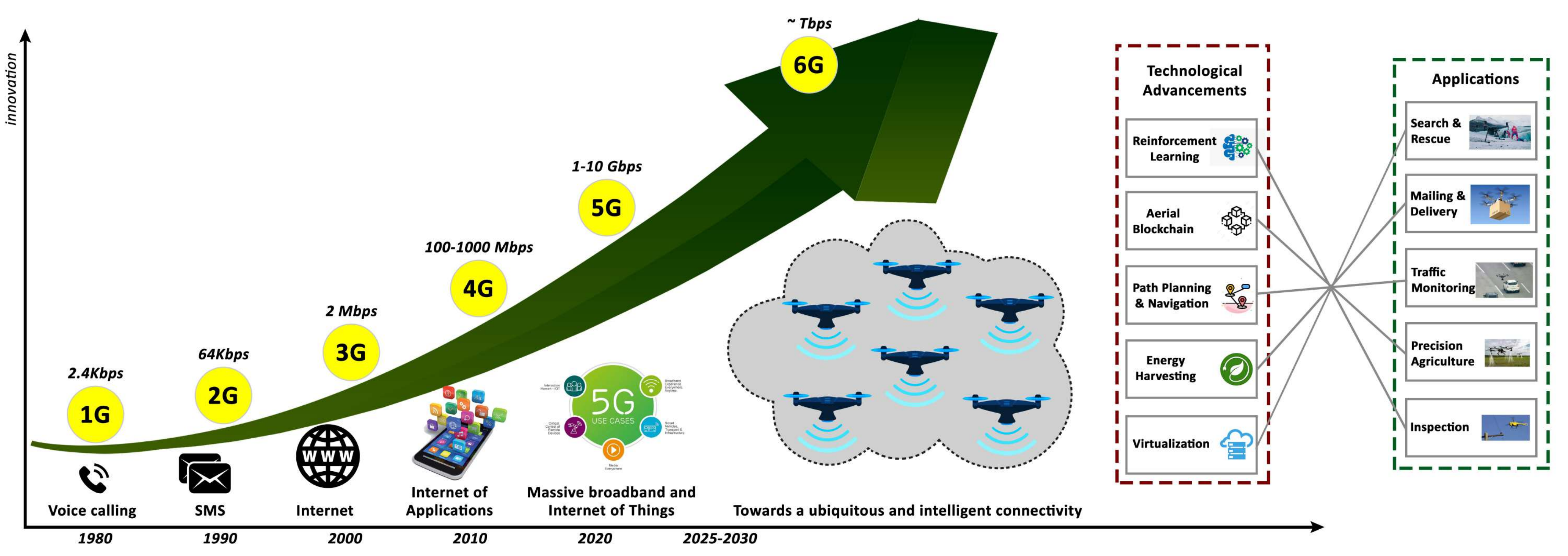}
    \caption{Evaluation of cellular communication with technological advancements for various applications of UAVs.}
    \label{fig:my_label}
\end{figure*}

\subsection{Existing literature and their limitations}
\label{sec:literature}

Over the past couple of years, a number of good reviews, tutorials and surveys articles focused on UAV communication networks over 5G/B5G have been published  \cite{35,36,37,38,39,40,41,42,3,44,45,46,47,48,49,50,51,52,53,54,55,56,57,58}. Table \ref{tableone} provides a brief summary of the recent and most popular articles in this domain. 
\begin{table*}[!htb]
\setlength{\bigstrutjot}{4pt}
\caption{1G to 6G – Key Performance Indicators (KPIs)}
\label{tabletwo}
\begin{adjustbox}{max width=1.0\textwidth}
\begin{tabular}{|l|L{1.5cm}|c|l|c|c|c|c|}
\cline{1-8}
 \multicolumn{1}{|c|}{\textbf{Network generation}}  & \textbf{Standards} &  \textbf{Core
Network} & \textbf{Frequency band} & \textbf{Mobility range} & \textbf{Theoretical data rate} & \textbf{Energy efficiency} & \textbf{Latency} \\  \cline{1-8}
\multirow{3}{*}{1G}  & \textbullet\ MTS & \multirow{3}{*}{PSTN}  &  \multirow{2}{*}{\textbullet\ 824-894
MHz} & \multirow{3}{*}{-} & \multirow{3}{*}{Up to 2.4 Kbps}  & \multirow{3}{*}{-} & \multirow{3}{*}{$>$ 1000 ms}  \\ 
& \textbullet AMPS &   & \multirow{2}{*} &  &   & &   \\
& \textbullet\ IMTS &    &  &   &   &  & \\
& \textbullet\ PTT &    &  &   &   &  &  \\ \hline 
\multirow{3}{*}{2G}  & \textbullet\ GSM & \multirow{3}{*}{PSTN}  & \multirow{2}{*}{\textbullet\ 850-1900 MHz} & \multirow{3}{*}{Up to 100 km/h} & \multirow{3}{*}{Up to 64 Kbps}  & \multirow{3}{*}{0.01 x} & \multirow{3}{*}{500 ms}  \\ 
& \textbullet\ IS-95 &   & \multirow{2}{*} &  &   & &   \\
& \textbullet\ CDMA &    &  &   &   &  &  \\ & \textbullet\ EDGE &    &  &   &   &  &  \\ \hline  
\multirow{3}{*}{3G}  & \textbullet\ UMTS & \multirow{3}{*}{Packet N/W}  & \multirow{2}{*}{\textbullet\ 1.8-2.5 GHz} & \multirow{3}{*}{Up to 150 km/h} & \multirow{3}{*}{Up to 2 Mbps}  & \multirow{3}{*}{0.1 x} & \multirow{3}{*}{100 ms}  \\ 
& \textbullet WCDMA &   & \multirow{2}{*} &  &   & &   \\ & \textbullet IMT2000 &    &  &   &   &  &  \\ & \textbullet CDMA2000 &    &  &   &   &  &  \\ & \textbullet TD-SCDMA &    &  &   &   &  &  \\ \hline
\multirow{3}{*}{4G}  & \  & \multirow{3}{*}{Internet}  & \multirow{2}{*}{\textbullet\ 2-8 GHz} & \multirow{3}{*}{Up to 350 km/h} & \multirow{3}{*}{Up to 1 Gbps}  & \multirow{3}{*}{1 x} & \multirow{3}{*}{50 ms}  \\ 
& \textbullet\ WiMAX &   &  &   &   &  &  \\ & \textbullet\ LTE  &   &  &   &   &  &  \\ & \textbullet\ LTE-A &    &  &   &   &  &  \\ \hline
\multirow{3}{*}{5G}  & \textbullet\ 5G NR & \multirow{3}{*}{IoT}  & \multirow{2}{*}{\textbullet\ Sub-6 GHz} & \multirow{3}{*}{Up to 500 km/h} & \multirow{3}{*}{Up to 10 Gbps}  & \multirow{3}{*}{10 x} & \multirow{3}{*}{5 ms}  \\ 
& \textbullet\ IPv6  &   & \multirow{2}{*}{\textbullet\ MmWave for fixed access} &  &   & &   \\
& \textbullet\ OFDM &    &  &   &   &  &  \\ \hline 
\multirow{4}{*}{6G}  & \textbullet\ GPS & \multirow{4}{*}{IoE}  & \multirow{2}{*}{\textbullet\ Sub-6 GHz} & \multirow{4}{*}{Up to 1000 km/h} & \multirow{4}{*}{Up to 1 Tbps}  & \multirow{4}{*}{$>$ 100 x} & \multirow{4}{*}{$<$ 1 ms}  \\ 
& \textbullet\ COMPASS &   & \multirow{2}{*}{\textbullet\ Exploration of THz bands (above 300 GHz)} &  &   & &   \\
& \textbullet\ GLONASS &    &  \multirow{2}{*}{\textbullet\ Non-RF (e.g., optical, VLC, etc.)} &   &   &  &  \\
& \textbullet\ Galileo &    &  &   &   &  &  \\ \hline 
\end{tabular}
\end{adjustbox}
\end{table*}
\begin{figure*}
\centering
    \includegraphics[width=0.9\textwidth]{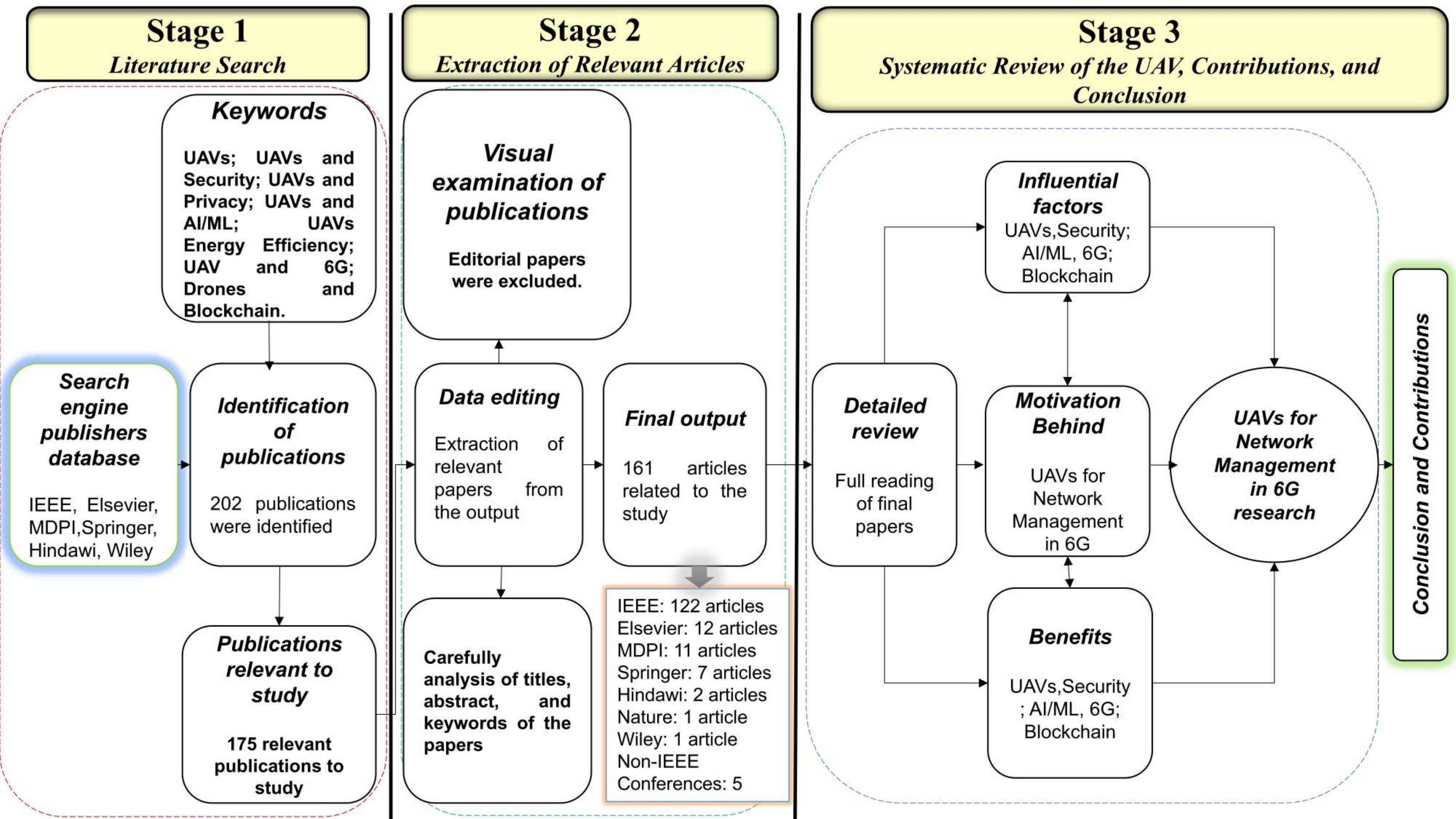}
    \caption{The methodology of the literature review and research process.}
    \label{fig:my_label}
\end{figure*}

More specifically, Li \textit{et al.} \cite{35} presented a comprehensive study of UAV communication over 5G/B5G wireless networks.The authors offered an overview of recent research activity on UAV communications incorporating 5G/B5G approaches from the viewpoints of the physical layer, network layer, cooperative communication, computation, and caching.The authors also investigated at certain open research topics in the hopes of building a solid basis for UAV applications in 5G/B5G wireless networks.Several challenges in UAV communication over beyond 5G wireless networks were discussed in a tutorial article presented by Zeng \textit{et al.} \cite{36}.Unique communication requirements and channel characteristics were among the highlighted challenges. Furthermore, key UAV network challenges such as energy limitation, high altitude, and rapid 3D mobility have also been investigated.

\begin{figure}
\centering
    \includegraphics[width=0.5\textwidth]{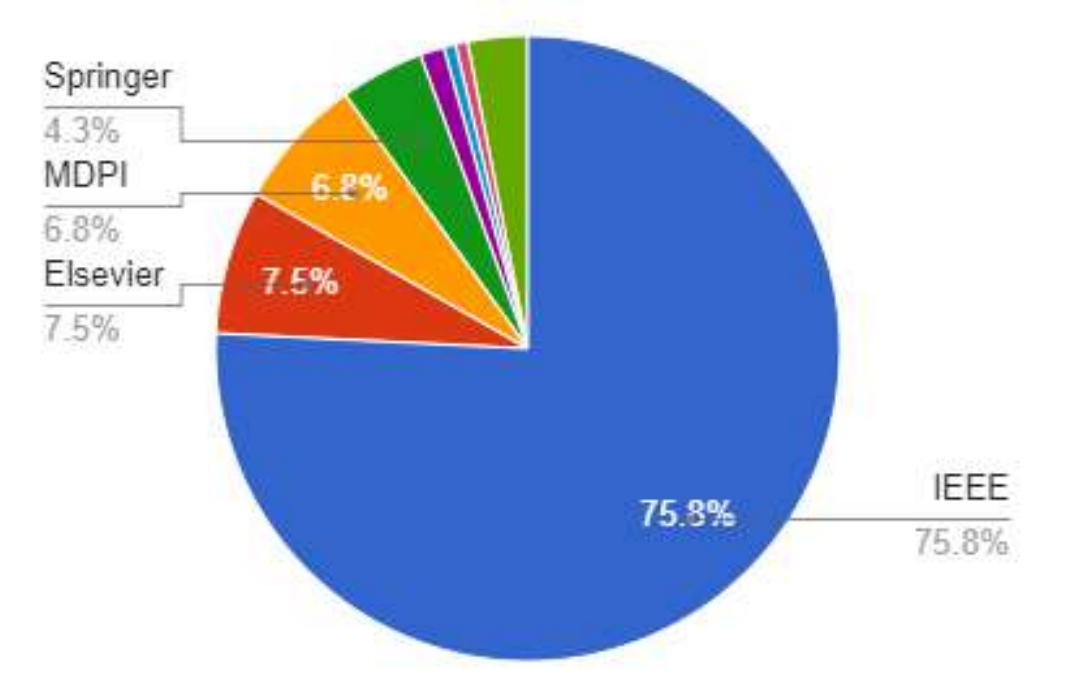}
    \caption{Percentage of articles from various publishers used in this review.}
    \label{fig:my_label}
\end{figure}

\begin{table*}[!htb]
\setlength{\bigstrutjot}{4pt}
\caption{Comparison with related reviews, tutorials and surveys articles focused on UAV communication networks over 5G/B5G and 6G. ($\surd$) shows that the topic has been covered. ($\times$) shows that the topic has not been covered. ($\partial$) shows that only a portion of the topic has been covered.}
\label{tableone}
\begin{adjustbox}{max width=1.0\textwidth}
\begin{tabular}{|l|L{1.1cm}|c|c|c|c|c|c|c|c|}
\cline{1-10}
 \multicolumn{1}{|c|}{\textbf{Year}}  & \textbf{Surveys} &  \textbf{Technology (5G, B5G,6G)} & \textbf{Security} & \textbf{Blockchain} & \textbf{AI/ML} & \textbf{Energy Efficiency} & \textbf{Comparison} & \textbf{Challenges} & \textbf{Open Research Topics on 6G} \\  \hline
 \multirow{2}{*}{2022}  & Ref. \cite{149} & $\surd$ & $\times$ & $\times$ & $\times$  & $\times$ & $\times$ & $\surd$ & $\surd$ \\ \cline{2-10}
& Ref. \cite{150} & $\surd$ & $\times$ & $\times$ & $\times$  & $\times$ & $\times$ & $\surd$ & $\surd$ \\ \hline 
\multirow{2}{*}{2021}  & Ref. \cite{47} & $\surd$ & $\surd$ & $\surd$ & $\times$  & $\times$ & $\surd$ & $\partial$ & $\surd$ \\ \cline{2-10}
& Ref. \cite{48} & $\surd$ & $\surd$ & $\surd$ & $\times$  & $\partial$ & $\times$ & $\surd$ & $\surd$ \\ 
\cline{2-10}
& Ref. \cite{170} & $\surd$ & $\times$ & $\times$ & $\times$  & $\partial$ & $\times$ & $\partial$ & $\surd$ \\ \hline 
\multirow{6}{*}{2020}  & Ref. \cite{39} & $\surd$ & $\times$ & $\times$ & $\surd$  & $\times$ & $\surd$ & $\surd$ & $\surd$ \\ \cline{2-10}
& Ref. \cite{40} & $\surd$  & $\surd$  & $\times$ & $\surd$  & $\times$ & $\surd$ & $\surd$ & $\surd$  \\ \cline{2-10}
& Ref. \cite{41} & $\surd$ & $\surd$ &  $\times$  & $\times$  & $\times$ & $\surd$ & $\surd$ & $\surd$ \\ \cline{2-10}
& Ref. \cite{3} & $\partial$ & $\surd$ &  $\times$  & $\partial$  & $\times$ & $\surd$ & $\surd$ & $\surd$ \\ \cline{2-10}
& Ref. \cite{46} & $\surd$ & $\partial$ &  $\partial$  & $\partial$  & $\times$ & $\times$ & $\surd$ & $\surd$ \\ \cline{2-10}
& Ref. \cite{45} & $\surd$ & $\times$ &  $\times$  &  $\surd$ & $\times$ & $\times$ & $\partial$ & $\surd$ \\ \hline
\multirow{3}{*}{2019} & Ref. \cite{35} & $\surd$ & $\surd$ &  $\times$  & $\times$  & $\times$ & $\surd$ & $\surd$ & $\surd$ \\ \cline{2-10}
& Ref. \cite{36} & $\surd$ & $\times$  & $\times$  & $\partial$  & $\times$ & $\times$ & $\partial$ & $\partial$\\ \cline{2-10}
& Ref. \cite{37} & $\surd$ &  $\surd$ & $\times$  & $\times$  & $\times$ & $\partial$ & $\surd$ & $\surd$\\ \hline
\multicolumn{2}{|c|}{Our survey} & $\surd$ & $\surd$ & $\surd$ & $\surd$ & $\surd$ &$\surd$ & $\surd$ & $\surd$ \\ \hline
\end{tabular}
\end{adjustbox}
\end{table*}

 Fotouhi \textit{et al.} \cite{37} published a survey that covered the majority of the factors that enable the smooth integration of UAVs into cellular networks. Future networks, such as 5G, are expected to be more equipped to deal with UAV-related issues. Sharma \textit{et al.} \cite{39} discussed recent advances in UAV communication and networking technologies. The paper examines UAV communication technologies as well as the use of centralized and decentralized techniques for both hardware and algorithm-based software. It was anticipated that the addition of 5G technology will provide a more stable and reliable networks. Ullah \textit{et al.} \cite{40} investigated the latest advancements in the integration of UAV networks into 5G and B5G systems. Standardization of UAVs, channel modelling, interference prevention, and collision avoidance were also investigated by the authors. Security and privacy issues, as well as optimal trajectory design employing deep reinforcement learning algorithms and energy harvesting approaches in UAV networks using 5G and B5G systems, are all thoroughly investigated.
 
Mishra \textit{et al.} \cite{41}  explored the complexities of integrating UAVs into 5G/B5G networks, as well as important technological developments in design prototyping and field testing, all of which supported the use of cellular-connected UAVs.  Alzahrani \textit{et al.} \cite{3} conducted a comprehensive evaluation of existing UAV-assisted research in areas such cellular communications, IoT networks, routing, data collecting, and disaster management. The authors also provided descriptions, classifications, and comparative evaluations of numerous UAV-assisted proposals. Zhang \textit{et al.} \cite{45} considered an Internet of UAVs over cellular networks, in which UAVs function as aerial users collecting different sensory data and transmitting it over cellular links to its transmission destinations. Noor \textit{et al.} \cite{46} presented a review article in which they explored key enabling technologies, applications, challenges, and open research topics for UAV networks in depth. According to the authors, the integration of 5G and 6G technologies will make UAV networks ultra-reliable and pervasive.
\begin{figure*}
\centering
    \includegraphics[width=0.9\textwidth]{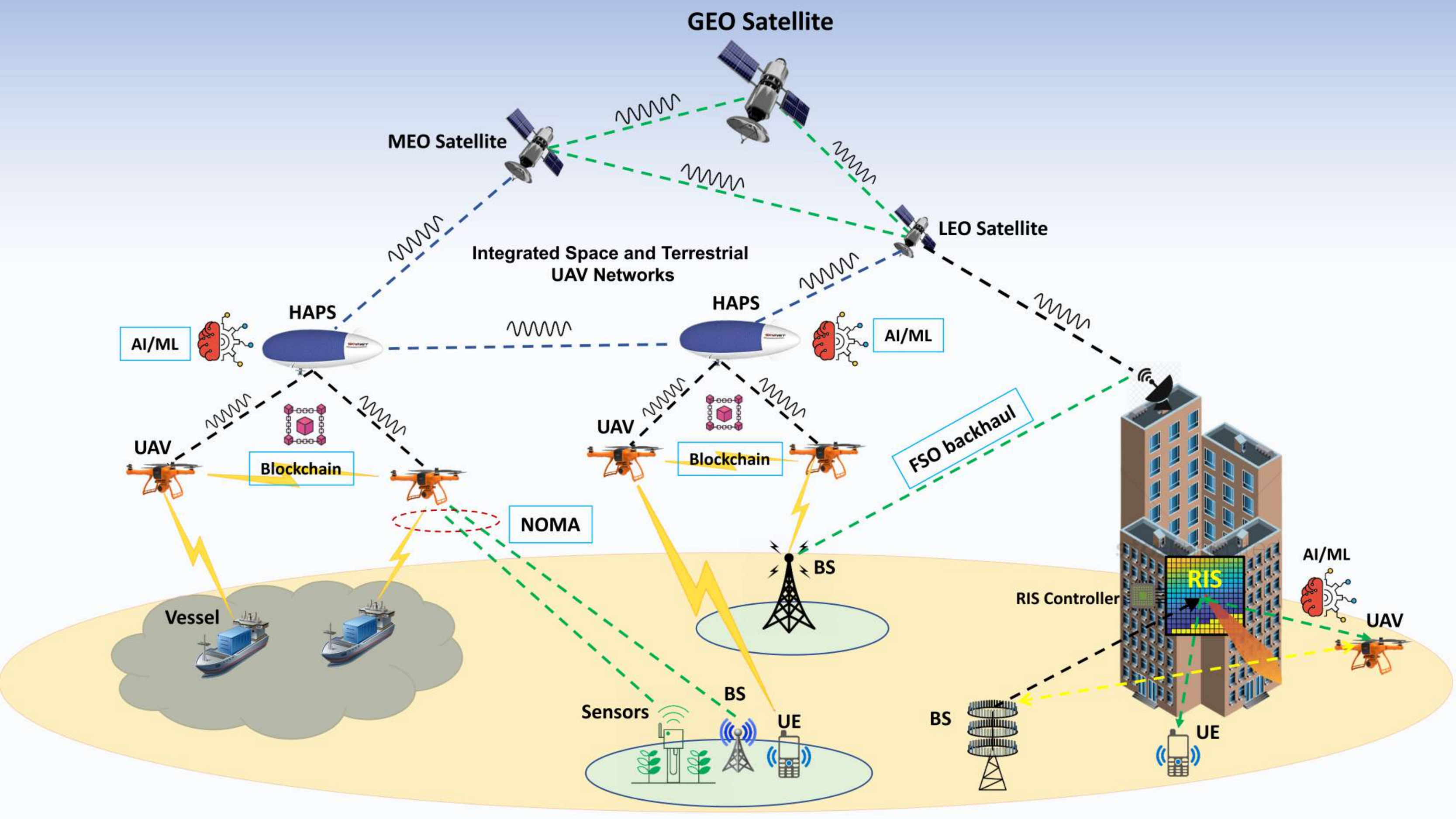}
    \caption{Integration of UAV networks with enabling technologies.}
    \label{fig:my_label}
\end{figure*}

Gupta \textit{et al.} \cite{47} demonstrated a blockchain-assisted secure UAV communication over 6G mobile networks. In the same article, research challenges and future directions for further improvement, as well as architecture, are also discussed. Han \textit{et al.} \cite{48} looked at 5G communication networks and mobile edge computing (MEC) as viable technologies for supporting UAV-enabled ecosystems and overcoming fundamental UAV network problems including limited computation, storage, and coverage. They also debated the 5G and MEC alternatives, outlining the latest advances and attempting to address some of the critical problems. Alongside that, following the recent popularity of UAV communication networks, they raised new security issues. The article also looked at contributions to the evolving drone industry that allow the use of each of the innovations listed. Wu \textit{et al.} \cite{170} provided a comprehensive overview of the recent research efforts on integrating UAVs into cellular networks, with a focus on exploiting advanced techniques such as intelligent reflecting surface, short packet transmission, energy harvesting, joint communication and radar sensing, and edge intelligence to meet the diverse service requirements of the next generation of wireless systems. In addition, the authors identified important directions for further investigation in future work.

 Bajracharya  \textit{et al.} \cite{149} presented a WI UAV with a 6G new radio running in the unlicensed band, which could be used as a relay, base station, or data collection/dissemination point. In this article, the authors have classed UAVs based on their characteristics, functions, and operations. Various regulatory and standardization efforts to integrate UAVs into the cellular network are being investigated. Several NR-U opportunities and design problems for WI UAVs are covered, as well as future scopes of WI UAVs. Azari \textit{et al.} \cite{150}recently analysed the potential prospects and use cases for THz-empowered UAV systems employing 6G networks, as well as the unique design limitations and trade-offs that go along with them. The authors also discussed recent developments in UAV deployment regulations, THz standardization, and THz-related health problems.
 
 \subsection{Research Methodology}
 Bibliographic research has been used to collect, analyze, and present the material obtained for review data, which included a variety of perspectives and methodologies. In this study, however, we employ a fundamental procedure\cite{169},to identify and filter the existing literature on the issue of UAV integration into 6G networks. Fig. 2 depicts the article screening procedure in details. In major databases, IEEE Explore, Web of Science, ScienceDirect, and MDPI, we used keywords, "UAVs", "UAVs and security", "UAVs and privacy", "6G," and other similar keywords, to identify possibly relevant publications. In the first round, 202 journal articles, conference proceedings, and early access publications were identified. Nonetheless, several irrelevant or unqualified items were eliminated. We used two criteria to exclude these items. We examined the titles, abstracts, and keywords of each of the chosen articles in great depth to confirm that each publication was really concerning UAVs and 6G mobile networks. Alternatively, we analyzed the papers using Scopus, Vosviewer, and Citavi, in that order. After removing extraneous papers, 161 relevant journal and conference proceedings publications were discovered.Fig.3 displays the percentage of articles from various publishers used in this review. Evaluating the existing literature allows for the identification of research gaps, hence highlighting research gaps.
 
\subsection{Contributions with Organization of the Article}

Almost all relevant articles in the literature have focused on 5G/B5G for UAV networks, but none of them covers all aspects of UAV networks over 6G. As a result, now is the time to present a detailed, up-to-date review article that addresses all aspects of UAV networks in a single go. To the best of our knowledge, this is the first review article covering all aspects of UAV networks. All of the top corners where we excelled are mentioned below:

\textit{\textbf{Integration of UAVs into 6G networks}\textit{}} (Section II): To understand the concept of integrating swarm of UAVs into 6G, we first present the network architecture, which is shown in Fig. 4. This comprehensive computing architecture is predicted to be the primary enabler for the vast majority of computationally intensive applications available on 6G systems. After evaluating the proposed architecture, we present a possible six-F trend in 6G mobile communications that could prove beneficial for UAV networks.

\textit{\textbf{Security Landscape}\textit{}} (Section III): We address the security issues that could prevent UAVs from being deployed in various 6G applications. We elaborate upon the essential security requirements that UAV networks must guaranteed for their successful functioning, including confidentiality, integrity, availability, authentication, trust, non-repudiation and authorization. The conventional security techniques for meeting these major security needs are discussed in depth. Finally, the necessity of offering lightweight cryptography approaches and the development of adaptive security tools are investigated.

\textit{\textbf{Blockchain Technology}\textit{}} (Section IV): The review also provides a larger perspective on how the integration of blockchain and 6G can address the security challenges of UAV networks.The fundamental blockchain characteristics of immutability, decentralization, and transparency that could be beneficial to UAV networks leveraging 6G mobile networks are highlighted.

\textit{\textbf{AI/ML Techniques}\textit{}} (Section V): To actualize UAV networks in 6G systems, we explain how 6G mobile networks can successfully support swarm of UAVs through the use of enhanced features and emerging AI and ML-based solutions. In addition, reinforcement learning (RL) techniques that can be used to find the optimal path to prevent collisions during real-time path planning and navigation missions are investigated.

\textit{\textbf{Energy Efficiency}\textit{}} (Section VI): The research explored UAV energy efficiency, which is a major stumbling block to their widespread adoption in a wide range of applications. As a general rule, flight time is the key problem for small UAVs because their onboard batteries have a finite lifespan. As a result, completing resource-intensive applications for UAV networks efficiently is a critical problem to be overcome.

\textit{\textbf{Challenges}\textit{}} (Section VII):
We examine the significant challenges and possible solutions to the integration of UAV networks into the 6G system as a means of advancing research in this area. UAV networks must overcome several challenges, including those related to safety, limited energy, storage and computation capability, routing, device compatibility, and high spectrum exploitation, as well as standardization and regulations.

\textit{\textbf{Open Research Topics}\textit{}} (Section VIII): In the context of 6G, we describe and explore open research topics for UAV networks.These open research topics will allow future UAV networks with 6G connectivity to reach their full potential.

\section{Integration of UAVs into 6G networks}

The sixth-generation (6G) mobile network can support a wide range of UAV network applications, including autonomous services and emerging trends.Ultra-high data density (uHDD), ultra-high speed low latency communications (uHSLLC), and ubiquitous mobile ultra-broadband (uMUB) are all 6G service classifications. The uMUB enables 6G systems to meet a wide range of performance needs in the space-aerial-terrestrial-sea segments. The uHSLLC provides ultrahigh data speeds and minimum latency, while the uHDD service class offers higher data density and reliability. For forthcoming uMUB, uHSLLC, and uHDD services, which are mostly absent in 5G and B5G networks, end-to-end codesign of communication, control, and computing elements will be required. 6G technologies are listed in Tab.IV for the uMUB, uHSLLC, mMTC, and uHDD services \cite {9}. One or more services can be improved by using each method. The following is a potential six-F trend in 6G mobile communications \cite {148}, which could be beneficial for UAV networks: 

\textit{\textbf{Full spectral:}\textit{} }A hyperspectral and full spectral system will be available for UAV networks in 6G, ranging from microwave, mm-wave, and terahertz to LASER.

\textit{\textbf{Full coverage:}\textit{}} UAV networks will provide full coverage in the terrestrial, aerial, space, and maritime domains using 6G mobile communications.

\textit{\textbf{Full dimension:}\textit{}} Holographic radio and communication will be entirely coherent.The 6G network will be exceedingly precise for UAV networks, allowing for accurate RF operation and a shift away from simple averaging and toward fine-grained analysis, modulation, and manipulation in the intensity-phase-frequency space.

\textit{\textbf{Full convergence:}\textit{}} The 6G network will be a multi-functional system that will compete with 5G's exclusive wireless communications capability. This multi-functional technology has the potential to introduce a surge of new revolutionary apps and services to UAV networks. All aspects of communication, control, sensing, computing, and imaging will be converged.

\textit{\textbf{Full photonics:}\textit{}} In 6G, photonics-defined radio (PDR) will be employed, together with a UTC photodetector (PD)-coupled antenna array, photonic engine, and spectrum computation. The adoption of full photonic processing will help UAV networks become more energy efficient.

\textit{\textit{\textbf{Full intelligence:}}} In the 6G mobile system for UAV networks, there will be ubiquitous and distributed computing and intelligence from the application layer to the physical layer.

6G could be helpful for UAV networks since it would allow UAVs to connect directly to one another while still being able to communicate with fixed infrastructure. Due to their limited on-board computing, storage, and battery life, UAV networks face several obstacles when it comes to accomplishing complex tasks successfully. However, it is feasible to overcome these obstacles by shifting computation- and storage-intensive tasks from resource-constrained UAVs to remote cloud servers exploiting the cloud computing capabilities enabled by 6G technology. Due to the fact that UAVs can be deployed as flying base stations (BSs) using physical layering techniques such as mmWave and massive MIMO, cognitive radios, and other technologies, data-intensive service needs can be satisfied \cite{59}. 

The most significant innovation in 6G will be satellite integration, which will allow UAV networks to provide centimeter-level precise positioning, global coverage, and heterogeneous QoS provisioning \cite{60}. Furthermore, combining satellites with 6G connectivity will provide a peak data throughput of 1 TBPS per device, as well as autonomous mobility of 1000 km/h in a highly populated urban setting. In the meantime, satellite operators are focused on a multi-layer airborne component system that incorporates the high altitude platform system (HAPS) and unmanned aerial vehicles (UAVs) to provide cost-effective communication services in hard-to-reach places. HAPS, which are usually located above the stratosphere, can give better coverage and cooperate with satellites to establish a more stable network, especially when satellite communications are hampered by bad weather. The proposed architecture of a 6G-enabled UAV network along with interactions among different technologies is illustrated in Fig.5.

Aside from the benefits of 6G for UAVs noted above, there are some unique challenges that UAVs likely encounter in future 6G networks \cite{61}. Typical UAVs can move at speeds of around 30–460 km/h while flying at various altitudes, resulting in a 3D mobility pattern. Base stations, on the other hand, are generally designed for ground coverage. In addition, antenna tilt can cause link fluctuation and coverage loss for UAVs in specific locations. As a result, the 6G network architecture should give a consistent signal at typical UAV flight altitudes of several hundred feet. Advanced antenna approaches like massive MIMO and adaptive 3D beamforming can help solve these challenges.

While autonomously executing missions, UAVs must be positioned and navigated to avoid colliding with one another and other objects such as buildings and trees. In many cases, the Global Navigation Satellite System (GNSS) delivers precise coordinates. However, for BLOS activities in particular, depending just on GNSS for navigation is inadequate. Using beamforming and triangularization from many base stations, 6G can provide additional positioning services. UAVs must be positioned and navigated to avoid colliding with one other and other objects such as buildings and trees while executing missions autonomously. In many cases, GNSS delivers precise coordinates. However, for BLOS activities in particular, depending just on GNSS for navigation is inadequate. Using beamforming and triangularization from many base stations, 6G can provide additional positioning services.
\begin{table}[h!]
\caption{Characterization of emerging technologies under different 6G services\cite{9}}
\label{tab:my-table}
\begin{tabular}{|l|l|l|l|l|}
\hline
Technology  & uMUB & uHSLLC & mMTC & uHDD \\ \hline
Unmanned aerial vehicles   &      $\surd$ &    $\surd$ &    $\surd$ &   $\surd$ \\ \hline
Artificial intelligence       &      $\surd$ &    $\surd$ &    $\surd$ &   $\surd$ \\ \hline
Terahertz communications       & $\surd$ &    $\surd$     &      &      \\ \hline
OWC                           &      $\surd$ &    $\surd$ &    $\surd$ &   $\surd$ \\ \hline
FSO backhaul/fronthaul       & $\surd$ &    $\surd$     &      &      \\ \hline
Blockchain                     &      &  $\surd$ &    &      \\ \hline
Massive MIMO                   &      & $\surd$ &    $\surd$ &   $\surd$ \\ \hline
3D networking 
 & $\surd$ &    $\surd$ &  &  $\surd$ \\ \hline

Quantum communications        
&     &  $\surd$ & $\surd$ & \\ \hline

Mobile edge computing          
&  $\surd$   &        &      &      \\ \hline
Backscatter communications   
&      &        &  $\surd$   &      \\ \hline
Intelligent reflecting surface  &      & $\surd$ &    $\surd$ &   $\surd$ \\ \hline
Dynamic network slicing      & $\surd$ &    $\surd$     &      &      \\ \hline
\end{tabular}
\end{table}

\subsection{Summary}
This section examined the integration of UAV networks into 6G. To illustrate this concept, we presented the network architecture depicted in Fig.4 at first. Future implementations of this integration could give the swarm of UAVs with rapid computing services, greater mobility, and increased scalability and availability. It will also make it easier to solve challenges like as security and privacy, limited processing, and battery resources, which are typically associated with UAVs.Satellite integration, the most major innovation of 6G, will enable UAV networks to provide centimeter-level precise positioning, global coverage, and heterogeneous QoS provisioning. In addition, the integration of satellites and 6G connection will allow a peak data throughput of 1 TBPS per device, as well as 1,000 km/h of autonomous mobility in inaccessible areas.

\section{Security landscape}

In this section, we discuss the security threats that could prevent UAVs from being employed in various 6G applications, as well as the security requirements for future deployment of UAV network.

\subsection{Security and Privacy Threats}

Due to design restrictions, small UAVs are not built with security and privacy threats in mind, leaving UAV networks exposed to both cyber and physical attacks \cite{62}. Intruders who aim to compromise the UAV network's security and privacy have a variety of options for carrying out their malicious intentions \cite{62}. They could, for example, send out a large number of reservation requests, eavesdrop in on control messages, and/or fabricate data exchange \cite{64}. Due to unreliable connections and inadequate security protocols, UAV networks linked to WiFi are known to be more vulnerable than cellular networks (i.e., 5G, B5G, and 6G) \cite{65}. Anybody with a right transmitter can bind to a UAV network and embed commands into an ongoing session, allowing them to be easily intercepted. Another concern regarding security and privacy in UAV networks is that if UAVs fly over a hostile environment, they could become a luring target for physical attacks \cite{66}. In such cases, the intruder can deceive the captured UAVs to get access to internal data through standard interfaces or ports.

Global positioning system (GPS) spoofing \cite{67,68,69} is another major security threat, which happens when an attacker manipulates GPS signals of the UAV. In this attack, an adversary sends false GPS signals to a planned UAV at a marginally higher power than the real GPS signals to trick the UAV into believing it is somewhere else. As a consequence, the intruder will use this technique to send the UAV to a predetermined location where it can be easily intercepted \cite{70}. UAV networks in 6G, on the other hand, can be linked with new wireless technologies such as visible light communications and quantum communications, which could lead to new security threats. To deal with such security threats, additional security mechanisms and countermeasures will be required. In the next subsection, we will present a brief overview of the security requirements and their level of impact for UAV networks in future 6G mobile networks.
\begin{figure}
\centering
    \includegraphics[width=0.5\textwidth]{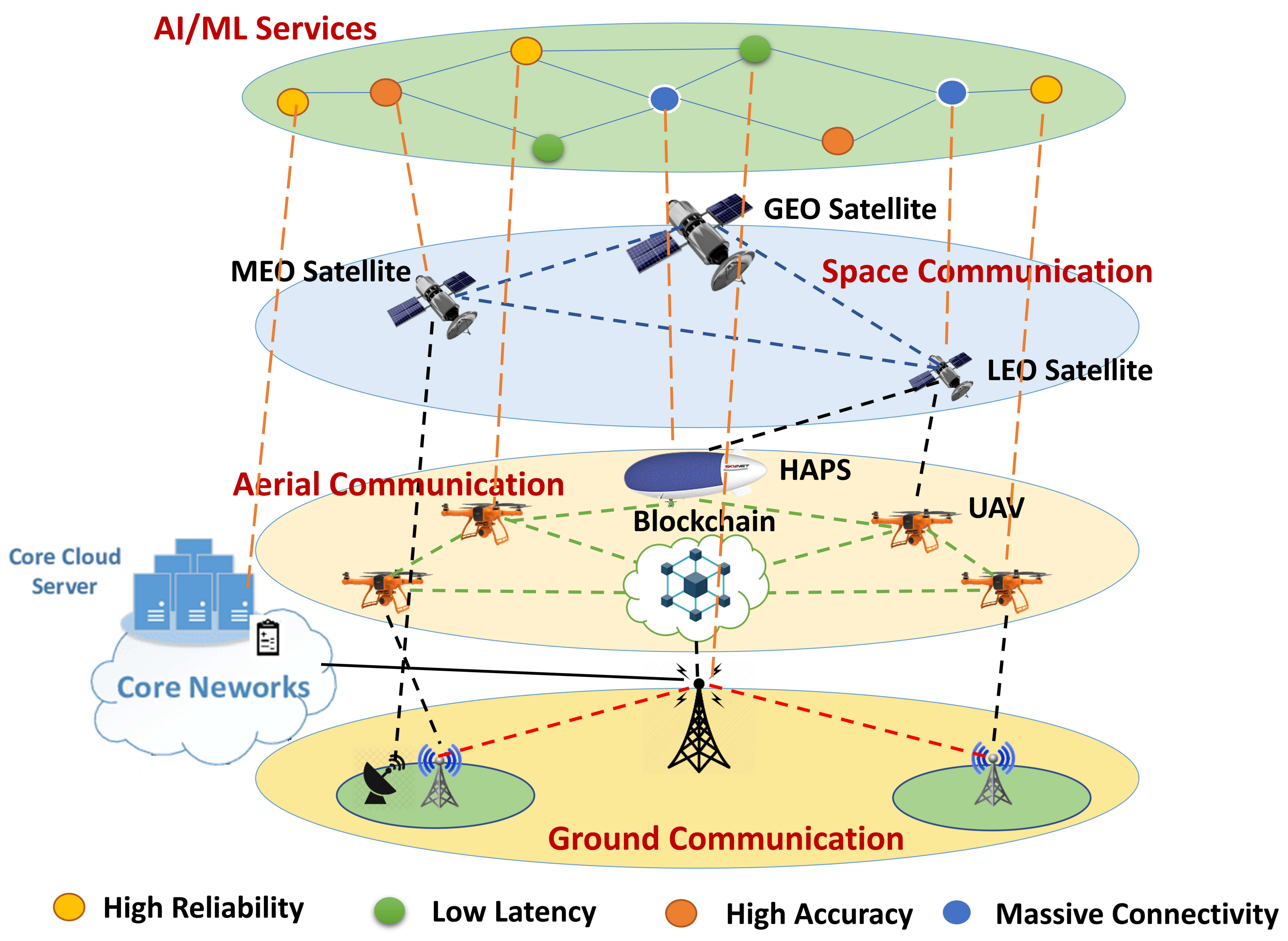}
    \caption{Proposed Architecture for 6G-enabled UAV Network with Interactions Among Different Technologies.}
    \label{fig:my_label}
\end{figure}

\subsection{Security and Privacy Requirements}

The widespread use of UAVs for a variety of civilian and commercial applications, as well as the ubiquitous wireless connectivity of future 6G networks, advanced security measures may be required to prevent unauthorized access to sensitive data \cite{71}. Likewise, while establishing security measures for UAV networks, characteristics such as high scalability, device diversity, and high mobility must be taken into consideration. Since 6G will include AI and Edge-AI based UAV functionalities, it is critical to ensure that security measures are in place to prevent against AI-related attacks. UAVs are very vulnerable to physical attacks due to their unmanned nature. By jamming control signals or using physical equipment, an adversary can physically capture UAVs and steal the crucial data they hold. Tab.V demonstrates the level of security requirements/impact for 6G-enabled UAV applications identified by the authors in \cite{72}. At the same time, security protocols must be designed with low communication and computation costs due to the restricted on-board computing capabilities of small UAVs.
To secure UAV networks against unauthorized access to sensitive information or other harmful attacks, the following key security and privacy properties must be guaranteed \cite{71,72,73}: 

\begin{itemize}
\item Confidentiality: In cryptography, confidentiality refers to ensuring that data is not made available or revealed to unauthorized users. Protecting sensitive data and data exchange between UAVs and the GCS from unauthorized access is critical in UAV networks because it could be a source of sensitive flight mission information leaks such as telemetry data and control commands. Encryption algorithms such as symmetric and asymmetric can be used to achieve confidentiality in UAV networks.
\item Integrity: Integrity refers to ensuring data consistency and trustworthiness during the communication process. Intruders may affect data integrity, which includes alterations such as modification, fabrication, substitutions, and data injections.  In UAV networks, data integrity is critical since it is a prerequisite for a successful flight mission. Hash algorithms with advanced encryption mechanisms can be employed to ensure data integrity.
\item Availability: The term "availability" refers to the fact that the services must be immediately available to authorized parties when they are needed for effective functioning. The goal of data accessibility is to ensure that legitimate users can get the information they need. Because the UAV network is utilized in mission-critical domains, the services must be available at all times without intentional or unintentional interruptions. Redundancy and backup may be useful for highly critical information services to ensure availability. Furthermore, the UAV system must be able to withstand classical denial-of-service (DoS) attacks that compromise its availability. Intrusion detection systems (IDS) can be used to resist such attacks.
\item Authentication: Authentication is a fundamental property that allows a UAV network to establish secure communication between their main components. It enables for the authentication and identification of UAVs taking part in the flight operation. The trustworthiness of each UAV is verified using a digital signature mechanism, and only authenticated UAVs are then allowed to participate in the flight mission. Authentication also protects the UAV network from adversaries who impersonate legitimate UAVs.  Another option for providing authentication in a UAV network is to use a blind signature scheme \cite{74}.
\item Trust: The term "trust" is defined as "confidence in an entity's integrity for the purpose of relying on it to perform particular tasks." Trust is dynamic, with different levels of assurance depending on particular criteria (such as identification, attestation, and non-repudiation) that determine when and how to rely on a connection. A UAV network connected to a 6G ecosystem will be characterized by a growing number of stakeholders and interconnected devices and services, not all of which will be managed by the same entity. Establishing trust in such an open and diversified environment is critical for the global adoption of this technology. Advanced cryptography schemes can be utilized to help implementing policies for achieving trust in UAV networks.
\item Non-Repudiation: In cryptography, non-repudiation is a property that prevents an entity from denying earlier agreements or activities (e.g., transmitting or receiving data).The sender of information receives confirmation of delivery and the recipient receives verification of the sender's identity when this property is being used, so neither side can subsequently challenge the data's processing. An entity in a UAV network will be unable to deny that it has previously communicated a message using this property. The UAV network must establish protocols to assure non-repudiation, which is accomplished through the adoption of a digital signature mechanism.
\item Authorization: Authorization is a security method for identifying a user’s privileges or access levels to system resources such as files, services, data, and application features. Data in the UAV network should only be accessible to permitted users. Unauthorized users are not allowed to communicate in any manner with the UAV network. Furthermore, UAV networks must specify, which resources are accessible to authorized users. To keep track of who has access to such resources, access control policies must be implemented.
\end{itemize}

Appropriate security protocols employing lightweight cryptography methods must be established to meet the aforementioned security and privacy requirements, allowing for efficient and secure communication between the various components of the UAV system. The research community, on the other hand, is still working on securing UAV communication channels while improving network capacity. UAV authentication can further secure the communication channel by preventing impersonation and replay attacks \cite{151,152}.The design of UAV access controls schemes, such as authorization and authentication mechanisms, remains a difficult research problem in the UAV networks. Indeed, any unauthenticated UAVs should not participate in flight missions to collect data from other UAVs in the network.Learning-enabled cyberattacks and massive data breaches are more likely in UAV networks employing 6G than traditional security concerns.Indeed, the possibility of using UAVs for malicious purposes grows as more intelligence is delegated to them in 6G networks. Adaptive security schemes can be implemented, in which security processes adapt to the threats landscape in real time and adjust their operation to detect and mitigate any threats. Deep learning-based techniques\cite{163,164,165,166}, such as GANs and GNNs can be used to improve UAV operation and management by making threat detection more robust, proactively analyzing the security status, and so making UAV operation and management more reliable and efficient.

\begin{table}[h!]
\begin{center}
\caption{Level of security requirements/impact for 6G-enabled UAV applications.}
\label{notations}
\begin{tabular}{lp{3.8cm}}
\toprule 
Security and Privacy Property & Requirements/Impact  \\
\midrule
Ultra Lightweight Security & Medium\\
Zero-touch Security & High \\
Domain Specific Security & Low\\
Energy Efficiency  & High\\
Computation Cost & High\\
High Privacy & Low\\
Proactive Security & Medium\\
Security via Edge & High\\
Limited Resources & High\\
Diversity of Devices & Medium\\
High Mobility & High\\
Physical Tempering & Medium\\
\bottomrule 
\end{tabular}
\end{center}
\end{table}

\subsection{Summary}
In this section, we explored the security and privacy issues that may arise as a result of the 6G requirements. Confidentiality, integrity, availability, authentication, trust, non-repudiation, and authorization are some of the fundamental security considerations that UAV networks must meet to operate effectively. Complex cryptographic operations for a UAV involved in a mission are difficult to execute due to the UAVs' typically limited on-board processing capacity.The conventional approaches to security are incapable of tackling this challenge. This section concludes with the suggestion that lightweight cryptographic algorithms, such as HECC, and adaptive security solutions are required.

\section{Blockchain Technology}

A blockchain is a collection of blocks connected by a cryptographic hash function \cite{75}. Apart from the genesis block, the hash value of each block is the hash value of the parent block. The blockchain, which is operated by a mechanism called consensus, which is a set of rules for assuring agreement between all participants as a blockchain ledger, is accessible to every member of the network. As a decentralized solution, blockchain technology provides a distributed computing paradigm for secure and adaptive protection of privacy preferences. It has the ability to prevent security breaches and ensure the integrity of data collected by UAVs \cite{47}.

 The implementation of blockchain with 6G communication in UAV networks will strengthen cyber security defenses. The fundamental characteristics of blockchain in terms of immutability, decentralization, transparency that could benefit UAV networks employing 5G and 6G mobile networks are as follows \cite{77}:

\begin{itemize}
\item Immutability: It refers to the ability of a blockchain ledger to stay unchanged and unmodified after being stored on the blockchain. This is because each block uses a hash function to link to other blocks, which is a one-way, irreversible process. The immutability property of blockchain technology can enable secure data storage and exchange in UAV networks using 6G mobile communication. Due to the rapid growth of UAVs, a secure and reliable database to store and exchange large data across the 6G wireless network is required. Because blockchain technology is immutable, it is seen to be a potential option for ensuring the security and privacy of data in UAV networks.
\item Decentralization: It is refers to the fact that the blockchain database is not managed by a particular entity or a central authority.Blockchain employs consensus algorithms such as proof of work (PoW) to build a secure chain of blocks and maintain the database's security without the usage of external control points. This significant element enables the development of a database platform with high immutability and robustness, as well as low data retrieval latency. This property will considerably lessen the impact of a single point of failure in UAV networks.
\item Transparency: All information about transactions on blockchain is visible to all entities in the network, which is called as transparency. To put it another way, a copy of the data records is replicated across the network of participants for public validation. Due to this feature, any entity can use its ability to check transactions based on its functions. This property will aid in improving node cooperation in UAV networks, hence improving data integrity. This feature is especially useful in 6G environments, where fairness and transparency are critical.
\end{itemize}

The above properties of blockchain can be utilized to improve the performance of the UAV-6G communication network by enabling secure spectrum sharing between network operators and UAV service providers. These features ensure privacy by utilizing a distributed information sharing platform and reducing the danger of hostile nodes abusing spectrum \cite{78}. A typical architecture for a blockchain-enabled 6G UAV network is depicted in Fig 6. As demonstrated in Fig. 6, a blockchain-enabled 6G UAV network can provide network security by utilizing an edge computing platform and core network.UAV networks can be useful in implementing MEC, especially in emergency situations like disasters where fixed ground infrastructure is unavailable. MEC is also viewed as one of the key technologies in 5G \cite {154}and expected to contribute towards 6G, and therefore it will require significant research to make it ultra-reliable in terms of survivability, availability, and connectivity.In addition, core network offers service adaptability, migration, collaboration, and evolution in UAV communication using AI as a function operating in the edge core network \cite {155}. The implementation of blockchain technology could allow participating UAVs to maintain high levels of trust and establish a flat architecture. Distributed entities generate transactions, which are subsequently recorded in blocks that are added to the blockchain. UAVs also periodically evaluate transactions in order to identify malicious nodes, offering trust and anonymity that protects the UAV network from malicious attacks. In addition, BC technology can secure confidential data collected by UAVs from unauthorised access by intruders.

Despite the potential benefits of blockchain in UAV networks, there are still a number of challenges that need to be addressed before blockchain can be implemented in the 6G mobile system.The majority of existing UAVs are resource-constrained in terms of computation. Cryptography and/or consensus mechanisms are frequently used in blockchain systems, however UAVs are usually incapable of doing computationally intensive tasks. Lightweight cryptographic solutions like Hyper Elliptic Curve (HEC) cryptography could be integrated with blockchain to solve this problem \cite {74,167}. A UAV network can allow a group of UAVs to execute a variety of tasks. The blockchain consensus of UAV networks can help to reduce the falsification of malicious UAVs and other security issues. Creating a scalable blockchain-based UAV network is challenging due to frequent topology change (i.e., UAVs can join and depart at any moment) and the scalability limitations of current blockchain systems. As a result, more study into the scalability of blockchain-based UAV networks is required in the future. In UAV networks, this challenge can be handled with efficient routing algorithms with blockchain using 6G technologies.

\subsection{Summary}
In this section, we discussed how the integration of blockchain and 6G could handle the security concerns of UAV networks. The core blockchain characteristics of immutability, decentralization, and transparency are noted as potentially advantageous to UAV networks employing 6G mobile networks.However, because to frequent topology changes and the scalability constraints of current blockchain systems, constructing a scalable blockchain-based UAV network is challenging. As a result, more research into the scalability of blockchain-based UAV networks is necessary. This problem can be solved in UAV networks by combining efficient routing algorithms with blockchain and 6G systems.
\begin{figure}
\centering
    \includegraphics[width=0.45\textwidth]{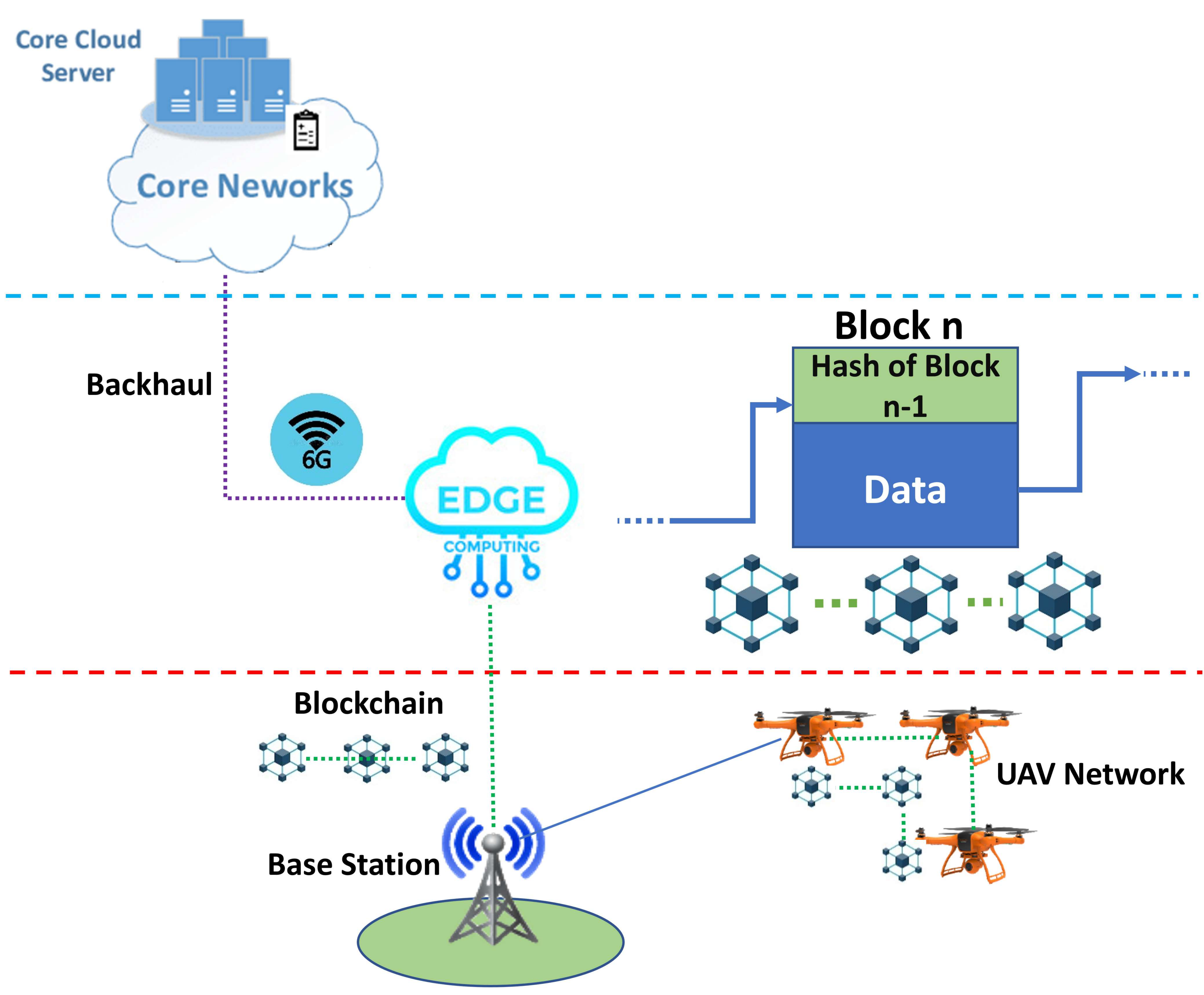}
    \caption{A Blockchain-enabled 6G UAV Network.}
    \label{fig:my_label}
\end{figure}

\section{AI/ML Techniques}

6G wireless networks will revolutionize the wireless evolution from "\textit{connected things}" to "\textit{connected intelligence}" \cite{79,156,157}. AI will play a critical role to guarantee the efficiency of future wireless communication networks, and it will represent the enabling technology for several applications \cite{80}. UAVs are one of these vital applications, which is expected to be a hot research area in the coming decades. AI, DL, and ML techniques are being applied to different aspects for the UAVs, which have shown prominent improvements in efficiency, resilience, and robustness \cite{81}.

There is a lot of literature linking AI/ML to UAV networks, and a few of them are summarized in tab.VI. Jung \textit{et al.} \cite{82} provided an interesting idea about a response-time to choose between processing data on-board or transmitting it using a MultiPath TCP (MPTCP) based on the artificial intelligence in order to increase performance of the network. In addition, Park \textit{et al.} \cite{83} have simulated the packet transmission rates of a UAVs using ML to computing the success and failure probabilities of transmission. This study recommended support vector machine with quadratic kernel technique, which shown that it faster and more accurate than linear regression technique based on the results provided. While, Khan \textit{et al.} \cite{84} proposed a new auto relay method based on UAVs for millimeter-wave communications to enhance the communication between ground and sky systems. In comparison to KNN and TR algorithms, directionality is adjusted via frequent matrix updates and real-time samples of link quality to determine ideal positions, resulting in improved stability and accuracy. Xiao \textit{et al.} \cite{85} proposed an RNN assisted framework to improve communication efficiency between a UAV and a BS due to the weather such as wind perturbation. Prediction air-to-air path loss and channel propagation are studied in \cite{86,87}. Zhang \textit{et al.} \cite{86} air-to-air path loss is investigated by using KNN and the Random Forest algorithms, and results are compared to empirical results. While, Alsamhi \textit{et al.} \cite{87} are used ANN to predict the signal strength of the UAV and estimate the channel propagation. However, these suggestions \cite{86,87} are considered not suitable for real-time application. 

Wang \textit{et al.} \cite{88} proposed a novel UAV communication paradigm in which UAVs can communicate using visible light and the communication is highly dependent on the ambient illumination. To optimise UAV deployment and minimise total transmit power, a machine learning algorithm that combines gated recurrent units (GRUs) and convolutional neural networks (CNNs) was utilised. While, Zhang \textit{et al.} \cite{89} investigated the optimal deployment of aerial BSs to offload terrestrial BSs by predicting the congestion in the wireless network to minimizing the power consumption of the drones. On the other hand, during the disasters cases, UAVs are consider a good solution of the WSN; Masroor \textit{et al.} \cite{90} are highlighted this issue and recommended that WSN with UAV together to increase the response of the network.

Aside from the work mentioned above, there are some tutorials and surveys oriented toward the use of machine learning methods in wireless communication networks. In \cite{159}, for example, offers a tutorial on artificial neural networks (ANNs) for wireless networks. All of the papers discussed above, however, do not expressly examine AI techniques for UAV applications.Furthermore, we noted that the majority of existing studies focuses on typical centralised methodologies for RL solutions, which presents a number of issues in terms of complexity and time management. That is why we believe distributed RL, such as the distributed Q-learning algorithm, is a promising method for solving real-time UAV applications. This form of RL approach is ideally suited for UAV networks where multi-agent choices must be made collaboratively\cite{158}.

We conclude this section by emphasizing that, because to the multi-dimensional nature of UAVs, adding AI/ML into UAV networks would surely present some challenges. As a result, one of the primary difficulties that researchers are continuously researching is selecting the optimal AI/ML methodologies. In addition, because to the intensive computations of AI/ML approaches, the improvement in latency between the UAV and the ground station should be addressed; as a result, researchers should work on increasing computational efficiency and optimizing performance to minimize latency. To provide a comprehensive upgrading of UAV networks, a large-scale deployment of AI/ML approaches at multiple network levels is necessary, which involves fundamental network changes. Furthermore, research into issues like as position verification, route management, and estimating the success rate of missions involving UAVs should be pursued in order to develop efficient and reliable AI-based UAV networks.Federated learning (FL), a potential distributed AI paradigm for collaboratively training a shared global model without revealing local sensing data, needs to be investigated in UAV networks \cite{160,161,162}.

\subsection{Summary}

In this section, we have shown how AI and ML-based solutions can enable 6G mobile networks to successfully support a swarm of UAVs, a necessary step toward implementing UAV networks in 6G systems. In addition, real-time path planning and navigation tasks that involve the avoidance of collisions are researched in order to better understand how reinforcement learning (RL) approaches can be employed to find the optimal path.Due to the multidimensional nature of UAVs, we highlighted that the adoption of AI/ML into UAV networks will probably provide challenges. While attempting to discover the optimal AI/ML methodologies, researchers confront a variety of obstacles. To decrease latency, it is vital to enhance computational efficiency. A comprehensive upgrading of UAV networks requires the widespread deployment of AI/ML methodologies across several network levels. To enhance this progress, research should be undertaken on areas such as location verification, route management, and estimating the success rate of UAV missions. AI-powered UAV networks that are trustworthy and effective to perform the real-time operation.

\begin{table*}[!htb]
\caption{AI/ML in UAV Networks}
\centering
\begin{tabular}{|p{2cm}|p{5cm}|p{9.5 cm}|}
\hline
Reference               & Key Contribution/Metric                             & Technique(s)                                                                                   \\ \hline
Jung et al. {[}80{]}    & Response-time between UAVs and BS                   & MultiPath TCP (MPTCP) based on the AI                                                          \\ \hline
Park et al. {[}81{]}    & Packet transmission rates of UAVs                   & Support Vector Machine (SVM)                                                                   \\ \hline
Khan et al. {[}82{]}    & Communication link between UAVs and BS systems      & K-Nearest Neighbors (KNN)                                                                      \\ \hline
Xiao et al. {[}83{]}    & Communication efficiency between a UAV and a BS     & Recurrent Neural Network (RNN)                                                                 \\ \hline
Zhang et al. {[}84{]}   & Air-to-air path loss                                & KNN and the Random Forest                                                                      \\ \hline
Alsamhi et al. {[}85{]} & Estimate the channel propagation                    & Artificial Neural Network (ANN)                                                                \\ \hline
Wang et al. {[}86{]}    & Optimize UAV deployment and minimize transmit power & Gated Recurrent Units (GRUs) and Convolutional Neural Networks (CNNs)                          \\ \hline
Zhang et al. {[}87{]}   & Optimal deployment of aerial BSs.                   & ML framework based on Gaussian Mixture Model (GMM) and Weighted Expectation Maximization (WEM) \\ \hline
Masroor et al. {[}88{]} & optimal deployment based on disasters cases         & Integer Linear Optimization Problem (ILP)                                                      \\ \hline
\end{tabular}
\end{table*}
\begin{figure}
\centering
    \includegraphics[width=0.45\textwidth]{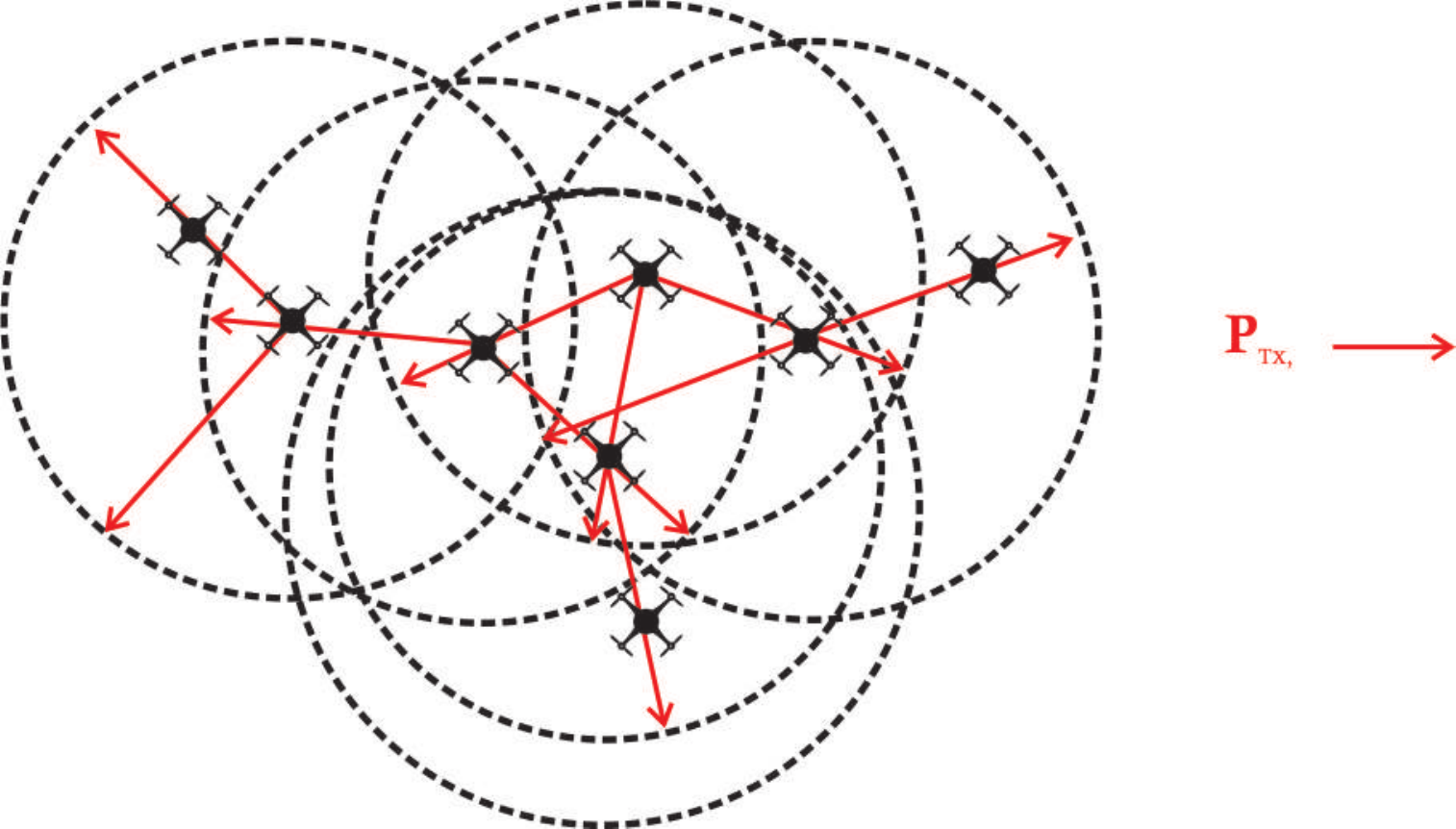}
    \caption{Maximum Transmission Range\cite{94}.}
    \label{fig:my_label}
\end{figure}
\section{Energy Efficiency }

Energy efficiency is defined as the ratio of the utility effect, whether in the form of a manufactured product or process, or simply the effect of a device, to the energy expenditure necessary to perform the planned activities. Improving energy efficiency is, to put it simply, a much more efficient use of energy to carry out the same process \cite{91,51,93}. The process of energy efficiency may concern many aspects of the functioning of the network as well \cite{94}.

A significant advantage of UAVs is a high degree of freedom, mobility in three dimensions \cite{95} and a relatively low cost of the device \cite{96}. An example is the expansion of the UAV network infrastructure as an access point (AP), which leads to the implementation of a scalable surveillance network with three-dimensional vision of the UAV \cite{98,99}. There are many different requirements for these networks to operate UAV applications well. it is important to pay attention to dynamic changes regarding topology \cite{100}, link failure \cite{101}, resource constraints \cite{102}, because the UAV network must be operated in an environment resistant to such changes \cite{103}.

A network that serves many UAVs must be fast and flexible. UAV network design and research has suffered from lack of application and many other reasons. of greatest importance in these studies is energy efficiency, concerning networks of unmanned aerial vehicles where there are energy leaks through communication. And this affects the throughput of the entire network \cite{104}. In the standard configuration of UAV network settings, one UAV sends messages with the same power level to all UAVs within transmission range. Building such complete connections ensures high network stability. however, this method of networking is ineffective \cite{105}. Then the UAV network also becomes ineffective by constantly generating more energy consumption than is actually needed \cite{106}.

Fig. 7 illustrates a UAV that uses the same transmitting power PTx to form a link with all other UAVs within its transmission range. A graphical depiction of the network topology, structured like an MST, is also shown in Fig. 8. The transmit power of the ith UAV, which is handled by a central or global controller, is referred to as PTx. The UAV network's root of the tree could be a gateway or a node sink \cite{107}. Although, because of the lower routing overhead, this centralised solution can dramatically cut power usage. Although power consumption per hop is lowered when all UAVs are connected by a few routes, total network connection becomes unstable owing to fewer routing options, which is crucial for highly mobile network UAVs. It's crucial to pay attention to the amount of hops, as this can be a role in higher energy usage \cite{108}.. 

\begin{figure}
\centering
    \includegraphics[width=0.45\textwidth]{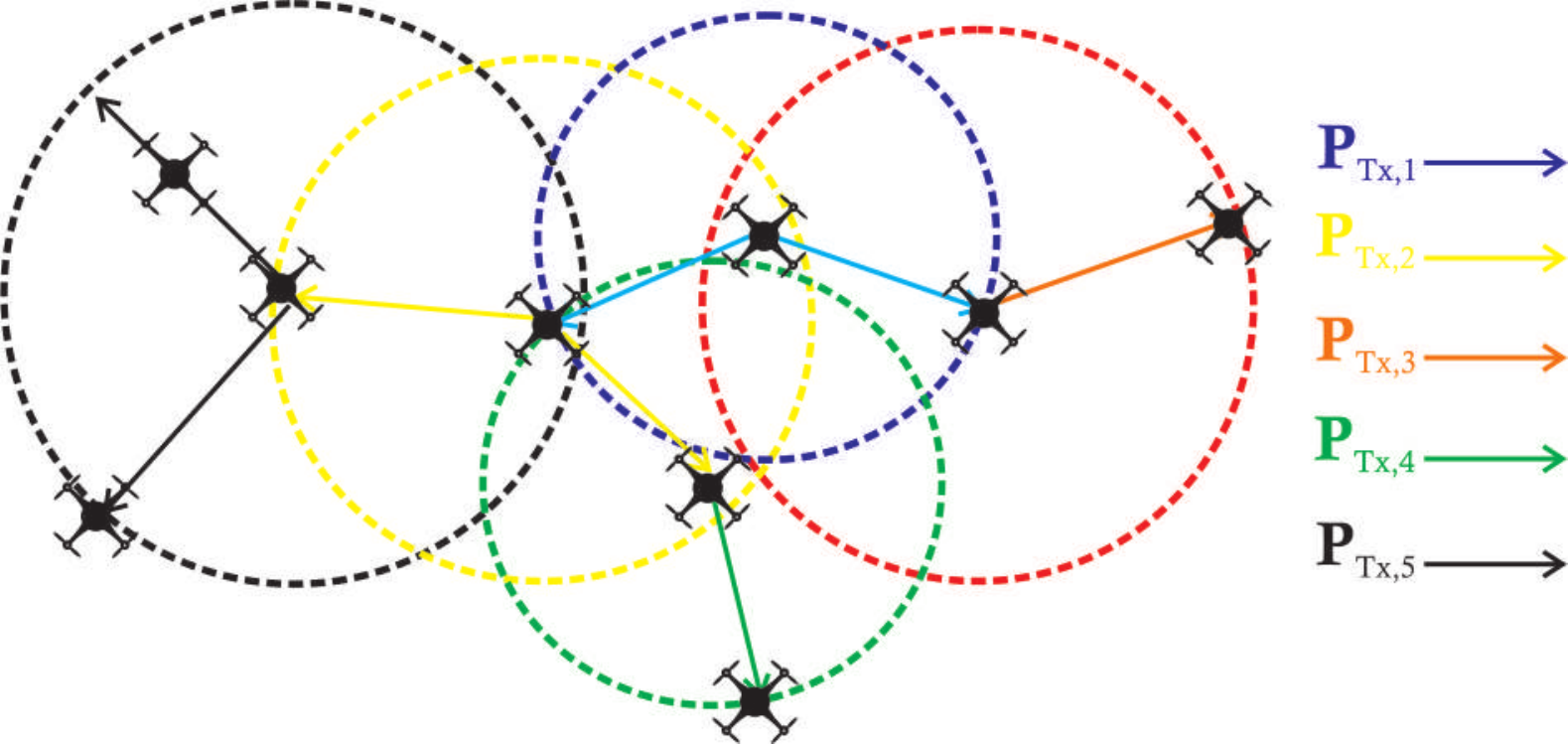}
    \caption{Centrally Controlled Transmission Range\cite{94}.}
    \label{fig:my_label}
\end{figure}

Distributed topology means that each UAV has a variable transmission power regulation \cite{109}. By clearly controlling the available UAV links, it can reduce energy consumption while maintaining the robustness of the network connection. With this network topology control method, Seongjoon Park \textit{et al.} \cite{94,95} proved that a UAV network can be formed from energy-saving network properties. The diagram of its functioning is shown in Fig.9.

\begin{figure}
\centering
    \includegraphics[width=0.45\textwidth]{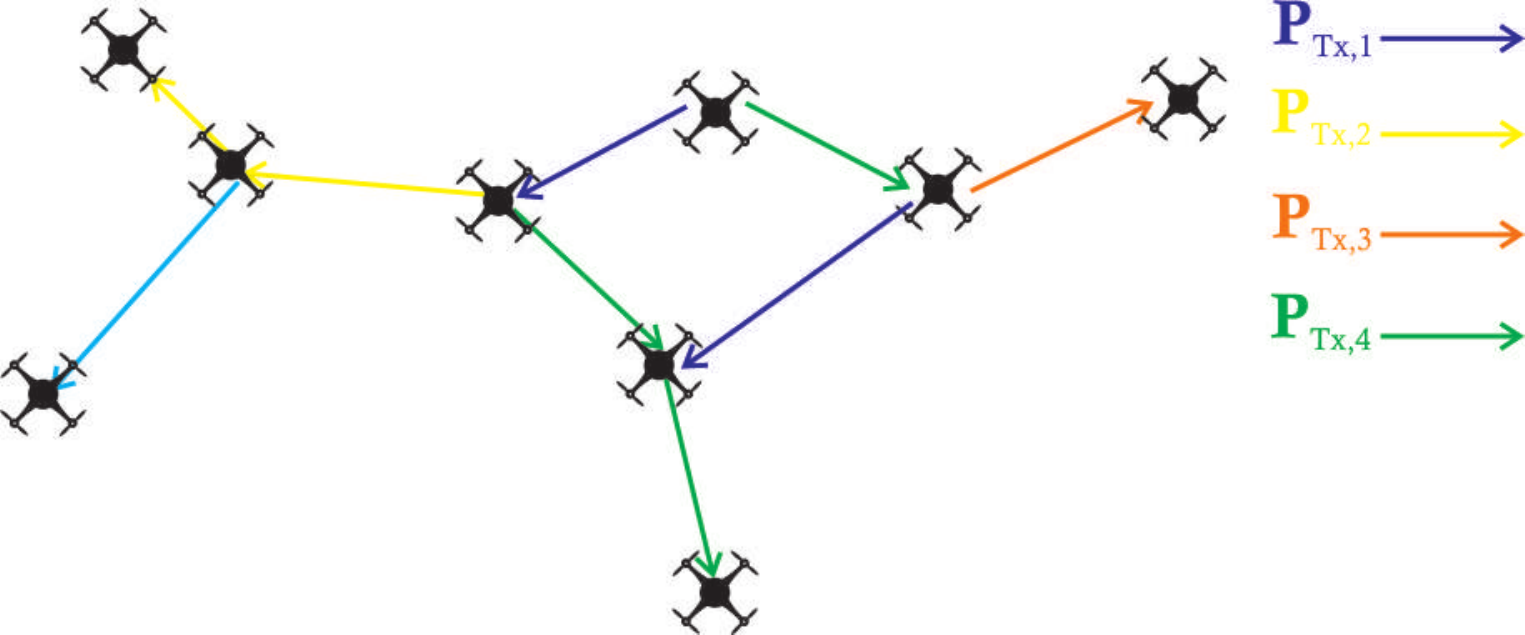}
    \caption{Example of topology control layer with 4 partitions\cite{94}.}
    \label{fig:my_label}
\end{figure}

The methodology of controlling the topology of energy-saving UAV networks has been described. The described system acts as an intermediary between the network and data links, so it is resistant to any other network environment. The space partitioning method is highly reducing energy consumption in end-to-end connections, while maintaining the node degree and the number of hops in the right amount.

Wang \textit{et al.} \cite{110}, as well as Noh \textit{et al.} \cite{111}, paid particular attention in their research to the energy efficiency of UAVs as base stations, but the energy consumption of UAV propulsion was omitted. In contrast, Hua  \textit{et al.} \cite{112} proposed to minimize the UAV's transmit power to achieve an energy-efficient deployment while providing wireless coverage for terrestrial users. Zeng and other researchers prove in their study that the energy efficiency of 5G systems assisted by UAVs with interference recognition was maximized when handling communication between devices \cite{113, 114}. The energy that consumes is the AUV drive is usually higher than that necessary for communication, the optimization of the drive energy consumption directly extends the operation time of the UAV \cite{51,115}. This relationship shows that the energy consumption for UAV propulsion cannot be disregarded. In a study by Duo \textit{et al.} \cite{100,116}, the authors investigated how to maximize the energy efficiency of a UAV based on a UAV-powered UAV model of energy consumption was used for safe communication. Furthermore, none of the prior researches consider the influence of UAV band allocation on energy efficiency. Hua et al. and other researchers looked into the subject of maximising UAV energy efficiency by working together to optimise user allocation, transmit power, bandwidth allocation, and UAV trajectory \cite{112}.

In addition to energy-efficient implementation, it is also vital to provide each user's needed QoS delay, according to certain prior scientific studies on networks supporting UAVs in degraded situations \cite{118}. It is extremely difficult and impracticable to achieve a deterministic delay in wireless networks supporting UAVs since the wireless channel has an inherent time-varying feature \cite{116}. To address the issue, effective bandwidth has been widely implemented in UAV networks to give users with statistical delay QoS. Hassan and his colleagues further increased the system's overall effective capacity by optimising the 3D UAV position and resource distribution while reducing the statistical QoS delay for each user \cite{118}.

Despite many different studies of combining UAVs with 5G and 6G techniques, research into UAV assisted wireless networks are still in the preliminary stage and many open problems require further or even in-depth research \cite{104,120}. Pay attention to interesting research topics for future directions, such as energy efficiency \cite{93,112,118}. Energy limitation is a bottleneck in any UAV communication scenario.

\subsection{Summary}
This section examined the energy efficiency of UAVs, which is a big hurdle to their widespread adoption for a variety of applications in future 6G mobile networks. Typically, flight duration is the most significant issue for small UAVs. However, unlike military UAVs, the vast majority of commercial UAVs are powered by an extremely limited capacity on-board battery. Most commercially available off-the-shelf UAVs can only fly for around thirty minutes. This restricts their use in applications that need long-running operations. Therefore, successfully executing resource-intensive applications for UAV networks is a critical concern that must be resolved. In this section, relevant literature on the issue of energy efficiency is discussed. In addition to the energy-efficient implementation of UAV networks, the provision of each user's required QoS delay in networks supporting UAVs in degraded environments has also been discussed.
\section{Challenges}
UAV communication and networking over 6G is in its early stages of growth. Since a UAV network will contribute to a range of applications in the 6G mobile network, there are several challenges that must be addressed for their deployed successfully, which we discuss in this section. To flexibly and securely integrate UAV networks into a 6G environment, as well as effectively allocating physical resources in UAV networks, these challenges require in-depth analysis.

\subsection{Safety}

The widespread use of UAVs for a variety of applications on 6G networks could raise severe safety concerns. UAVs that crash while completing their tasks can cause significant harm to both public property and human life. This could be the result of a technical failure, insufficient system service, mid-flight collisions, or operator error \cite{121}. Extreme weather conditions, such as turbulence, lighting, battery capacity limitations, and inadequate lifting capabilities, create concerns of UAVs collapsing over public property. Furthermore, since commercial planes share airspace in big cities, there is a significant risk of airborne collisions resulting in massive destruction.

\subsection{Limited Energy, Storage and Computation Capability}

Due to the minimal on-board resources such as battery, storage, and computation of small UAVs, integrating a UAV network with 6G may be challenging. Small UAVs rely on these on-board facilities in the general domain. It is not practicable to change UAV batteries in the air during a flight. Completing the resource-intensive applications on schedule is thus a critical problem. Moreover, the data gathered by small UAVs may be too large for a single UAV to process and store on-board while performing a monitoring task simultaneously \cite{122}. It demands a significant amount of processing and storage capability. As a result, undertaking computationally complex activities can cause UAVs to respond more slowly, reducing their efficiency.

\subsection{Routing}

Routing allows UAVs to communicate and collaborate with one another, as well as choose the optimal path for data transmission. Routing is the most challenging issue in a UAV network because of the unique characteristics of UAVs, such as high mobility, 3D movement, and frequent topological changes \cite{123}. For extremely sensitive applications, information interchange between the UAVs and the ground station must be reliable, stable, and efficient. However, in order to make applications and services more successful and active on 6G networks, suitable routing protocols for UAV communication must be designed and selected.

\subsection{Device Compatibility}

With the compatibility of 6G mobile networks, the resource-constrained nature of UAVs will be the most difficult challenge. For stand-alone UAVs, supporting 1 Tbps throughput, AI, XR, and integrated sensing with communication characteristics is difficult \cite{124}. Moreover, developing an on-board wireless module capable of transmitting and receiving mm-Wave for UAVs would be a hard process. Furthermore, integrating the mentioned technologies to improve the technical capabilities of UAVs to be compatible with 6G may result in greater expenses. 

\subsection{ High Spectrum Exploitation}

6G networks will extend even further, leveraging a larger and higher spectrum to enable Tbps connections, with THz and visible light communications being the most promising options \cite{125}. Despite advances in the literature in terms of exceptionally high spectrum utilization for intra-tier and inter-tier communications, effective and optimal communications will always be a challenge and a future study area for the scientific community. Because, to begin with, high-bandwidth communications are susceptible to attenuation due to changes in ambient conditions, particularly when UAVs move in three dimensions. Second, in order to improve transmission efficiency, super-narrow beamforming techniques with high directional degrees and quick transformation should be researched.

\subsection{Standardization and Regulations}

Terahertz spectrum allocation and usage regulations are a challenge in 6G mobile networks since they involve the coordination of different governments and locations throughout the world in order to allocate a standard band range as far as possible \cite{126}. In addition, the standards and rules regulating satellite communications would be a big challenge in a 6G mobile networks.  First and foremost, all governments must consult on satellite communications orbit and spectrum resources. Furthermore, when the UAV network extends into a range of vertical industries with drastically different characteristics, it will have to compete with significantly different user behavior. It will be a difficult challenge to shift users' natural ways of thinking and habits in these many vertical industries, and to adapt to new ways of thinking and standards of conduct as quickly as feasible.

\section{Open Research Topics}

Building on the identified challenges in Section VII, we now put forward and discusses the open research topics to spur further investigation of UAVs network in 6G contexts (summarized in Table III).
\subsection{Blockchain-enabled UAV Softwarization}

Blockchain technology can ensure the privacy and integrity of data gathered by UAVs \cite{127}. Similarly, by integrating blockchain and 6G mobile connectivity, UAV networks will be more secure against cyber-security vulnerabilities. Despite current study into blockchain technology for UAV networks, researchers have yet to investigate a blockchain-enabled softwarization for UAV networks. A blockchain-enabled softwarization for UAV networks could be used to offer 6G communication services with dynamic, adaptive, and on-the-fly decision capabilities \cite{128}. However, real-time deployment of highly mobile UAVs remains a difficulty. As a result, in order to meet privacy and security issues, real-time deployment is essential. Another problem is a single-point failure, where a centralised SDN controller controls all decisions for the whole UAV network.

\subsection{High-speed Backhaul FSO Connectivity }

The provision of a super high-speed, cost-effective, easy-to-deploy, and scalable backhaul link is essential in managing the massive quantity of data for linking UAV networks and the core network in 6G wireless networks. Free space optics (FSO) networks are a promising option for high-speed connectivity and to address the bottleneck problem in the backhaul link. Weather conditions, on the other hand, could have a significant impact on the vertical FSO link. One possible solution to this challenge is to design an adaptive algorithm that adjusts the transmit power and divergence angle in response to weather conditions. In rainy conditions, for example, high power and a small divergence angle could be adopted \cite{129}. Other options include using hybrid solutions, such as using millimeter-wave (mm-wave) spectrum in combination with FSO. Unlike FSO, fog has no effect on mm-waves. On the other hand, mm-waves are greatly attenuated by water molecules \cite{130}. UAV networks may use FSO in rainy conditions and switch to mm-waves in foggy weather. In bad weather, the hybrid FSO/RF connection is also a potential option for solving link degradation problems \cite{131}.

\subsection{Intrusion Detection Systems and Forensics Models}

An intrusion detection system (IDS) is required to identify intrusions against UAV networks during a flight mission in real-time, as well as forensics models to analyze the compromised UAVs in the event of an incident \cite{71}. Because UAV networks are a complex cyber-physical system that will include multiple components in 6G networks, intrusion detection and forensics methods should take into account various information gathering sources to improve performance. Taking into account several information sources, on the other hand, might raise communication and computation costs. Due to existing trade-offs between security and efficiency, developing such solutions is challenging. As a result, lightweight IDSs are required to monitor UAV networks and identify threats. Furthermore, forensics investigation of UAVs is a research topic in UAV security that has yet to be investigated. Existing digital forensics models do not have the necessary unification and standards to cover a larger range of commercial UAVs.

\subsection{6G Protocol Designs for UAV Networks }

As 6G evolves, new dynamic multiple access protocols will be required that could dynamically vary the type of multiple access (orthogonal or non-orthogonal, random or scheduled) employed according on the applications' demands and the network state \cite{17}. In compared to current 2D networks, the addition of new dimension such as altitude will significantly alter the connectivity nature. Therefore, novel handover protocols must be designed to account for the UAVs' 3D networking characteristics in the 6G system. In addition, all 6G protocols for UAV networks must be distributed and able to use data-sets deployed across the network edge. Furthermore, AI-driven signaling, scheduling, and coordination protocols are needed to replace conventional 5G protocols that rely on pre-determined network characteristics and arbitrary frame structures. 

\subsection{Intelligent Mobility Management}

In 5G and pre-5G mobility management, node movement was not taken into account. Fast-moving UAVs and their supporting satellites will add to the complexity of 6G mobility scenarios. Inter-UAV and inter-satellite connections alter when the UAV and satellite locations change. The network topology, handover control mechanisms, and other elements would be affected by the high mobility \cite{132, 168}. Intelligent mobility management should address different types of handovers for terminals with ongoing service connections, such as handover between beams, handover between UAVs and satellites and handover between UAVs and base stations. To ensure stable communication, deep reinforcement learning algorithms could be employed to dynamically adjust handover decisions. Deep reinforcement learning algorithms can be used to discover the optimal path to prevent collisions during real-time path planning and navigation.

\subsection{Reconfigurable Intelligent Surfaces}

Reconfigurable intelligent surfaces (RISs) are considered a promising technology for the 6G wireless networks that intelligently reflects incoming signals to improve coverage and capacity while enabling massive connectivity \cite{133}. RISs have also been shown to be a low-cost, green, sustainable, and energy-efficient option for 6G networks on a system level perspective \cite{134}.The use of RIS carried by UAVs to support cellular communications networks and services offers significant promise for expanding wireless communications and addressing the growing complexity of the wireless channel, as well as improving channel quality in urban areas \cite{135}. However, RIS-assisted UAVs network research is still in its infancy, and there are several opportunities for significant contributions and advancements in this sector. For example, the UAV's on-board battery and the weight, size, and number of RIS elements, which restrict the UAV's flying time, could be configured together to serve a specific mission. The RIS elements will also regulate the channel conditions in order to give the best signal routes for the outgoing links. UAVs' free mobility patterns will allow for quick and accurate alignment of the light emitting diodes and the RIS, ensuring LoS.

\subsection{Energy Harvesting Technologies}

A typical UAV's limited flight time is due to its insufficient battery capacity and payload capability, which is still a major barrier to their use in a wide range of applications in 6G wireless networks. The use of energy harvesting technologies to charge UAVs can overcome this problem\cite{136,137}. Solar-cell technology has recently improved to the point that the energy sources are now efficient enough to consider adding weight to a UAV. In addition, when there is no irradiation and therefore no power can be harvested from solar panels, a hybrid solar-RF energy harvesting system can be implemented on UAVs for powering the day and night flight. Because hybrid harvesting systems are more efficient than stand-alone harvesting systems, UAV flight duration could be greatly increased at any time of day.

\subsection{SDN and  NFV for UAV-enabled 6G Networks}

UAV networks have recently integrated the concepts of SDN and NFV to address their performance issues. SDN and NFV can help to reduce network management complexity \cite{138} and the need to deploy particular network devices for UAV integration \cite{139}. SDN can also be used to connect different VNFs.  To enable new IoT applications, UAV networks can be connected to the Internet utilising 6G technology by using cloud computing, web technologies, and service-oriented architectures \cite{140}. For such a networked environment, UAV resources can be virtualized with other network resources. As a result, effective solutions for virtualizing UAV-enabled 6G networks will need to be developed in the future.

\subsection{Integrated Space and Terrestrial UAV Networks}
To date, short-range communication technologies and cellular systems have been widely used to link UAV networks, and they rely heavily on terrestrial base stations. Satellites have long been the most common communication solution for oceanic, mountainous, and wild terrestrial locations where conventional ground communication networks are impractical or extremely costly to provide communication services \cite{141,147}.The deployment of non-terrestrial infrastructures as part of the 6G network, known as the integrated space and terrestrial UAV networks, is being considered as an emerging topic with the intent of enhancing coverage rates. Meanwhile, satellite operators are working on a multi-layer airborne component system that includes the HAPS and UAVs to enable cost-effective worldwide communication services. UAVs have evolved into a critical facilitator and vertical component of the future 6G ecosystem due to their low cost and great performance. However, a number of scientific challenges must be addressed, including the aerial platform's power supply, the antenna array's stability, channel models, seamless handover, admission control, interference management, and so on. 3D channel modelling \cite{142}, advanced multi-antenna technologies \cite{143}, spectrum-awareness, dynamic spectrum management, and FSO \cite{144} are some of the future research areas in this emerging field.

\subsection{Quantum Backscatter Communications}
Quantum backscatter communications (QBC)  \cite{145,146}, another promising technology, which will aid in the development of UAV networks and their integration with 6G, particularly in terms of QoS and security. In the QBC paradigm, the transmitter produces entangled photon pairs termed signal photons and idler photons. A backscatter transmitter emits and backscatters the signal photon, while the idler photon is retained at the receiver. The QBC configuration improves the communication channel's error exponent significantly and enables secure communication through quantum cryptography \cite{153}.Traditional security methods such as encryption and digital signatures may not be viable for a swarm of UAVs due to power and complexity constraints. On the other hand, this technology is not intended for large-scale deployment of low-power UAV networks, which often include a diverse set of devices and is one of the primary goals of 6G mobile networks. Multiple access technologies (for example, NOMA and rate splitting multiple access) have been identified as one of the promising choices for enabling massive connection in such networks while preserving high energy and spectrum efficiency.

\section{Conclusion}

In this paper, we provided an in-depth review of UAV communication and networking over 6G. The key barriers to widespread commercial deployment of UAV networks in 6G wireless systems, as well as potential solutions, have been outlined. With the integration of 6G mobile connectivity, UAV networks will become ultra-reliable and ubiquitous. However, security and privacy concerns, as well as limited on-board energy and processing capabilities, limit the use of UAVs in a variety of applications. A perfect balance of communication technologies, security schemes, intelligence, and energy harvesting methods is required to develop secure and efficient UAV networks with extended flight duration and minimal communication latency. As a result, this paper addressed the security and privacy, AI/ML, and energy-efficiency solutions that surround UAV networks in 6G. We provides an overview of how the 6G wireless network can be used to connect blockchain and AI/ML with UAV networks. The findings of the review are then given, along with possible challenges. In addition, we ended our review by shedding new light on future research prospects in this rapidly developing field. Despite the numerous challenges that we will confront on the way to 6G, the future of connected skies with UAV networks seems promising. Perhaps now is the time for researchers to consider interplanetary networking and communications for swarms of UAVs, as well as their integration with 6G.

\bibliography{mybib}
\bibliographystyle{IEEEtran}
\end{document}